\documentclass[final,5p,times,twocolumn,authoryear]{elsarticle}
\usepackage{aas_macros}
\usepackage{amssymb}
\usepackage{amsmath}
\usepackage{lipsum}
\usepackage[utf8]{inputenc}
\usepackage[T1]{fontenc}
\usepackage{hyperref}
\usepackage{xcolor, subcaption}
\usepackage{comment}
\usepackage{textcomp}
\usepackage{eurosym}
\usepackage{comment}

\newcommand{\ray}{\texttt{C$^2$-Ray}}
\newcommand{\pyray}{\texttt{pyC$^2$Ray}}
\newcommand{\hii}{{{H\fontsize{8}{12}\selectfont II}}}

\defcitealias{c2ray}{M06}

\journal{Astronomy $\&$ Computing}

\begin{document}
\begin{frontmatter}
\title{\pyray{}: A flexible and GPU-accelerated Radiative Transfer Framework\\ for Simulating the Cosmic Epoch of Reionization}

\author[first]{Patrick Hirling}
\author[first,third]{Michele Bianco\corref{corrauth}}
\ead{mb2158@sussex.ac.uk}
\cortext[corrauth]{Corresponding author}
\author[second]{Sambit K. Giri}
\author[third]{Ilian T. Iliev}
\author[fourth]{Garrelt Mellema}
\author[first]{Jean-Paul Kneib}
\affiliation[first]{organization={Institute of Physics, Laboratory of Astrophysics, Ecole Polytechnique Fédérale de Lausanne (EPFL)},
            addressline={Observatoire de Sauverny}, 
            city={Versoix},
            postcode={1290}, 
            country={Switzerland}}
\affiliation[second]{organization={Nordita, KTH Royal Institute of Technology and Stockholm University},
            addressline={Hannes Alfvéns väg 12}, 
            city={Stockholm},
            postcode={SE-106 91}, 
            country={Sweden}}
\affiliation[third]{organization={Astronomy Centre, Department of Physics \& Astronomy},
            addressline={Pevensey III Building, University of Sussex}, 
            city={Falmer, Brighton},
            postcode={BN1 9QH}, 
            country={United Kingdom}}
\affiliation[fourth]{organization={The Oskar Klein Centre, Department of Astronomy,},
            addressline={Stockholm University, AlbaNova}, 
            city={Stockholm},
            postcode={SE-10691}, 
            country={Sweden}}

\begin{abstract}
Detailed modeling of the evolution of neutral hydrogen in the intergalactic medium during the Epoch of Reionization, $5 \leq z \leq 20$, is critical in interpreting the cosmological signals from current and upcoming 21-cm experiments such as the Low-Frequency Array (LOFAR) and the Square Kilometre Array (SKA). Numerical radiative transfer codes provide the most physically accurate models of the reionization process. However, they are computationally expensive as they must encompass enormous cosmological volumes while accurately capturing astrophysical processes occurring at small scales ($\lesssim\rm Mpc$). Here, we present \pyray{}, an updated version of the massively parallel ray-tracing and chemistry code, \ray{}, which has been extensively employed in reionization simulations. The most time-consuming part of the code is calculating the hydrogen column density along the path of the ionizing photons. Here, we present the Accelerated Short-characteristics Octahedral ray-tracing (\texttt{ASORA}) method, a ray-tracing algorithm specifically designed to run on graphical processing units (GPUs). We include a modern \texttt{Python} interface, allowing easy and customized use of the code without compromising computational efficiency. We test \pyray{} on a series of standard ray-tracing tests and a complete cosmological simulation with volume size $(349\,\rm Mpc)^3$, mesh size of $250^3$ and approximately $10^6$ sources. Compared to the original code, \pyray{} achieves the same results with negligible fractional differences, $\sim 10^{-5}$, and a speedup factor of two orders of magnitude. Benchmark analysis shows that \texttt{ASORA} takes a few nanoseconds per source per voxel and scales linearly for an increasing number of sources and voxels within the ray-tracing radii.
\end{abstract}


\end{frontmatter}

\section{Introduction}\label{introduction}
The Epoch of Reionization (EoR) is a period of significant interest in the history of the Universe, as it marks the appearance of the very first sources of radiation that drove the transition of the intergalactic medium (IGM) from its primordial cold and neutral state to the present-day hot and highly ionized one \citep[see e.g.][for reviews about this era]{Furlanetto2006, Gorbunov2011book, Dayal2018early}. While indirect observational evidence, such as using high redshift quasar spectra \citep[e.g.][]{bosman2022hydrogen} and the cosmic microwave background (CMB) radiation \citep[e.g.][]{aghanim2020planck}, situates the EoR at redshifts between about 5 and 30, its main characteristics are still unknown \citep{pritchard201221, barkana2016rise}. Current and upcoming interferometric radio telescopes, such as the Low-Frequency Array \citep[LOFAR;][]{van2013lofar}, Hydrogen Epoch of Reionization Array \citep[HERA;][]{deboer2017hera}, Murchison Widefield Array \citep[MWA;][]{wayth2018mwa} and Square Kilometre Array \citep[SKA;][]{mellema2013ska}, are expected to uncover the details of this key event in cosmic history by detecting the distribution of the redshifted 21-cm signal in the IGM, produced by the spin-flip transitions in neutral hydrogen \citep{pritchard201221, Zaroubi2013EoR}. Accurate modeling of the EoR, which is needed to interpret the observational constraints provided by these experiments, will require performing detailed numerical radiative transfer (RT) and radiation hydrodynamics (RHD) studies on large cosmological scales ($\gtrsim 100$ Mpc). These simulations are challenging because the EoR is a non-local process, and the underlying RT equation contains both angular, spatial, and frequency dimensions. Various modeling methods exist, a review of which may be found in, e.g., \cite{gnedin2022modeling}.

Today, most fully numerical RT codes can be divided into two main classes: moment-based and ray-tracing methods. The former works by considering the hierarchy of angular moments of the RT equation, with some `closure relation' to limit the number of equations to be solved, and treat the radiation as a fluid \citep[e.g.][]{aubert2008radiative}. This makes coupling to hydrodynamics natural and, from a computational perspective, has the huge benefit of being independent of the number of ionizing sources in the simulation. On the other hand, moment methods suffer from increased diffusion and unrealistic shadows on optically thick objects. A few examples of codes using moment-based methods are \texttt{OTVET} \citep{otvet}, \texttt{RAMSES-RT} \citep{Ramses-RT} and \texttt{AREPO-RT} \citep{kannan2019arepo}. Although they combine N-body, hydrodynamic and radiative feedback, they tend to be computationally expensive and cannot simulate the required large volumes. \cite{Iliev2014Howbig} show that we require simulations with a minimum volume size of $\sim$100 cMpc to model the 21-cm signal and cover the large field of view expected by SKA and its precursors\citep[e.g.][]{mertens2020improved, trott2020deep, hera2023improved}. Moreover, the small mass sources ($\sim$10$^8 M_\odot$), which are expected to drive reionization \citep[e.g.][]{nebrin2023starbursts, gelli2023quiescent, atek2024most}, are often not resolved by these simulations.

Ray-tracing methods take a more physical approach by casting \emph{rays} around each source and modeling how the radiation propagates, i.e., is absorbed and scattered, along those rays. The photo-ionization rate occurring at any point in space is then determined by the number of absorptions between the source and said point, normally expressed as the optical depth between the source and that point. This approach can potentially be more accurate and less diffusive than moment methods but is quite expensive, as the cost of ray-tracing generally scales linearly with the number of radiating sources. Thus, in practice, the number of sources that can be considered has been, until recent years, severely limited by the available computational power. \ray{} \citep{c2ray}, \texttt{ZEUS-MP} \citep{zeus}, \texttt{CRASH} \citep{Ciardi2001CRASH}, \texttt{SPHRAY} \citep{Altay2008sphray}, \texttt{LICORNE} \citep{Semelin2007licorne}, \texttt{ART} \citep{Nakamoto2001art}, \texttt{FLASH-HC} \citep{Rijkhorst2006} are a few notable examples of ray-tracing-based codes. Moment and ray-tracing methods have been compared extensively \citep{comparison_1, comparison_2}. The main differences are due to numerical diffusion for the different treatments of the energy equation in moment-based methods and how the multi-frequency radiation is implemented. The advantages of using one over the other have been shown to depend greatly on the problem and context. That being said, by requiring huge volumes and large numbers of ionizing sources \citep{kaur2020minimum, giri2023suppressing}, developing more efficient RT methods for EoR, especially ray-tracing-based ones, is highly desirable.

In recent years, there has been a significant surge in the use of general-purpose GPUs for numerical scientific research. These devices have enabled remarkable performance improvements when used to develop applications for problems that can be divided into numerous simple and independent tasks suitable for parallel processing. Consequently, GPU acceleration has been integrated in various astrophysics and cosmological-related tools \citep[e.g.][]{Ocvirk2016CoDa, Potter2016pkdgrav3, Racz2019StePs, sph-exa, Wang2021Photons}. The \texttt{ATON} \citep{Aubert2010GPU} and \texttt{EMMA} \citep{Aubert2015emma} codes are the first applications of GPU-accelerated algorithms for radiative transfer codes in the context of extra-galactic astrophysics. To our knowledge, this technology has not yet been imported to short-characteristic ray-tracing methods, thus making the current work a first.

Given the success of GPUs in accelerating ray-tracing tasks in computer graphics \citep{Owens2008GPUcomput, Nickolls2010GPUera, navarro_hitschfeld-kahler_mateu_2014}, it is reasonable to explore their application to ray-tracing problems in astrophysics. This motivates our work, where we introduce an Accelerated Short-characteristics Octhaedral ray-tracing (\texttt{ASORA}) method designed specifically for \ray{}. 
By incorporating GPU methods, we anticipate significant performance enhancements and more efficient simulations, thus opening up new possibilities for research and analysis. Our work aims to bridge the gap between the potential of GPU acceleration and the requirements of ray-tracing tasks in astrophysics, providing a promising avenue for further advancements in this domain.
 
\ray{} is a 3D ray-tracing radiative transfer code designed for simulating the EoR and was initially developed by \cite{c2ray} (hereafter: \citetalias{c2ray}). It conserves photons at a voxel-by-voxel level, allowing for large, optically thick grid voxels while maintaining accuracy. Furthermore, the method allows for long time steps, even surpassing the voxel-crossing time of ionization fronts. It has been extensively used in EoR simulations and updated to include photoheating, X-ray radiation, and helium chemistry \citep{helium, ross2017simulating,ross2019evaluating}. \ray{} is written in \texttt{Fortran90} and designed for massively parallel systems, utilizing a hybrid MPI and OpenMP approach for efficient radiation propagation. The ionizing sources are distributed over MPI processes, and each of these processes further employs OpenMP threading to propagate radiation in a domain-decomposed manner.

As a stand-alone code, \ray{} is a \emph{post-processing} code\footnote{The algorithm can, however, be used in conjunction with a hydrodynamics code, as was, for example, done in \cite{Henney2011} and \cite{Medina2014}}. It acts on the output snapshots of a previous (cosmological) hydrodynamical simulation and propagates radiation on the gas fields of these snapshots. As is detailed in the following sections, it is also a grid code, meaning that the gas fields must be projected onto this grid through some smoothing method. Sources are identified in the initial simulation via a variety of models. Currently, the typical approach is to run a halo finder on each snapshot and use a physical model to translate a halo into a radiating source. The update to \ray{} in this work comprises two main aspects:
\begin{enumerate}
    \item \textbf{GPU-Accelerated ray-tracing Method}: The original ray-tracing method used by \ray{} is not well-suited for GPU parallelization. A new algorithm based on the same short-characteristics scheme has been developed to address this limitation. This new method is specifically designed for running on GPUs, enabling efficient computation of column densities, which is the most computationally intensive task in the radiative transfer (RT) method. The GPU implementation leverages massive multi-threading capabilities, resulting in significantly faster performance than the CPU method. This new algorithm is written as a C++/CUDA \citep[e.g.][]{garland2008CUDAparallel} library with \texttt{Python} bindings for ease of use and integration.
    
    \item \textbf{Python Wrapper and Interface}: The highly-optimized \texttt{Fortran90} implementation of \ray{} excels at computationally intensive and time-consuming tasks, such as the solving of chemistry equations and, until now, ray-tracing. However, due to its compiled and statically typed nature, \texttt{Fortran} is less suited for all the parts of the code that require frequent tweaking, such as the radiation source implementation, interfacing, I/O operations, cosmological model, and more generally the setup of each particular simulation. These tasks contain most of the conceptual baggage of future simulations but only represent a negligible fraction of the computational workload. Thus, to enhance usability and flexibility, we decided to wrap the time-critical core \texttt{Fortran} subroutines of \ray{} and rewrite the non-time-consuming parts of the code in \texttt{Python}, making frequent use of standard libraries.
\end{enumerate} 
As a result, users can now write an entire \ray{} simulation as a \texttt{Python} script, making it easier to tweak parameters and add new features without frequently recompiling the core \texttt{Fortran} subroutines. These updates enable more efficient GPU utilization for critical computations and improve the overall accessibility and versatility of the \ray{} code through \texttt{Python} scripting and interface enhancements.

This paper is structured as follows. In \S~(\ref{sec:c2ray}), we describe how reionization is modeled and summarize how the \ray{} method works. In \S~(\ref{sec:ray-tracing}), we describe the ray-tracing method used, present our newly developed \texttt{ASORA} algorithm, and briefly discuss the new \texttt{Python} wrapping and interface to the code. Then, in \S~(\ref{sec:results}), the updated code is tested on standard idealized situations and benchmarked to determine how much performance improvement is achieved. The source code of \pyray{} is publicly available at \url{https://github.com/cosmic-reionization/pyC2Ray}.

\section{Simulating Cosmic Reionization}\label{sec:c2ray}
To study the EoR, we need to model the time evolution of the ionization state of the intergalactic medium (IGM) within a cosmological framework. This involves solving a system of chemistry equations that track the evolution of the ionization state of primordial species, such as hydrogen and helium. These equations take into account various physical processes, including photoionization, collisional excitation, recombination, heating, and cooling \citep[e.g.][]{Furlanetto2006}.

In this paper, we will focus on the simplest case, considering only hydrogen. This choice is justified because hydrogen constitutes the major part of the IGM. The original \ray{} code includes extensions also to consider helium ionization and multi-frequency photo-heating \citep{friedrich2012radiative}, and we plan to incorporate these extensions into \pyray{} gradually. The primary objective of this paper is to present an update to the general ray-tracing method.

The ionization state of the hydrogen gas is described by the following \emph{chemistry equation} \citep[e.g.][]{Choudhury2006physics, Choudhury2009analytical},
\begin{equation}
    \frac{dx_\mathrm{HII}}{dt} = (1-x_\mathrm{HII})\left(\Gamma + n_e\,C_{\rm H}(T)\right) - x_\mathrm{HII}\,n_e\,\alpha_{\rm H}(T),
    \label{eq:chemistry}
\end{equation}
where $x_\mathrm{HII}$ is the fraction of ionized hydrogen, $n_e$ is the electron number density, $\Gamma$ is the photo-ionization rate per unit time, and $C_{\rm H}(T)$ and $\alpha_{\rm H}(T)$ are the collisional ionization and recombination coefficients for ionized hydrogen and free electrons, at temperature $T$.
\ray{} uses the on-the-spot (OTS) approximation, which assumes that the diffused photons resulting from recombination to the ground state are reabsorbed locally and, thus, solely accounted for by using a different value for $\alpha_{\rm H}$ \citep[e.g.,][]{Ritzerveld_ots}.

The photo-ionization rate $\Gamma$ quantifies the effect of ionizing UV radiation on the gas and is determined by the distribution of radiation sources. To illustrate this point, consider the simple situation of a single isotropic ionizing source in a homogeneous medium. As photons propagate away from the source in all directions, they form a spherical "shell" of ionizing radiation. The photons are absorbed by gas particles, which subsequently become ionized. These photo-ionizations also attenuate the strength of the radiation further away from the source, in addition to the attenuation occurring due to geometrical effects alone. Photo-ionization is also countered by recombinations. Together, these phenomena result in the formation of a spherical ionized bubble around the source, also known as a Strömgren sphere.

In EoR simulations, more than one source is typically present, and the medium is distinctly \emph{in}homogeneous, leading to a much more complicated situation. The number of these sources during the EoR depends critically on the size of the volume and minimum mass of source haloes. We typically start with less than a few hundred sources at high redshift ($z\gtrsim 20$) to a few tens of a million at low redshift ($z\approx 6$). Below, we first summarize the method used in \ray{} to solve \autoref{eq:chemistry}, and then discuss in detail the computation of the photoionization rates.

\subsection{Summary of the C$^2$Ray Method}\label{sec:c2raysummary}
To solve the chemistry equation (\autoref{eq:chemistry}), one could, in principle, use a finite-differencing scheme and assume all rates to be constant over a reasonably short timestep. This approach is used by, e.g., \texttt{Grackle} \citep{Grackle2017} to solve very complex chemistry networks. The problem here lies in the photoionization rate $\Gamma$. It is determined by the amount of ionizing photons arriving at the target point where \autoref{eq:chemistry} is considered. This amount depends directly on how radiation is absorbed along the path from its source to the target point, which in turn depends on the density $n_\mathrm{H}$ and ionization state $x_\mathrm{HII}$ of the medium along this path. This means that $\Gamma$ is strongly dependent on the solution variables of the problem and that this dependence is also highly non-local. For the finite-differencing scheme to be accurate, this implies very stringent constraints on the timestep size, especially in the presence of fast-moving ionization fronts (I-fronts).

\begin{figure}[t]
\centering
    \includegraphics[width=0.9\linewidth]{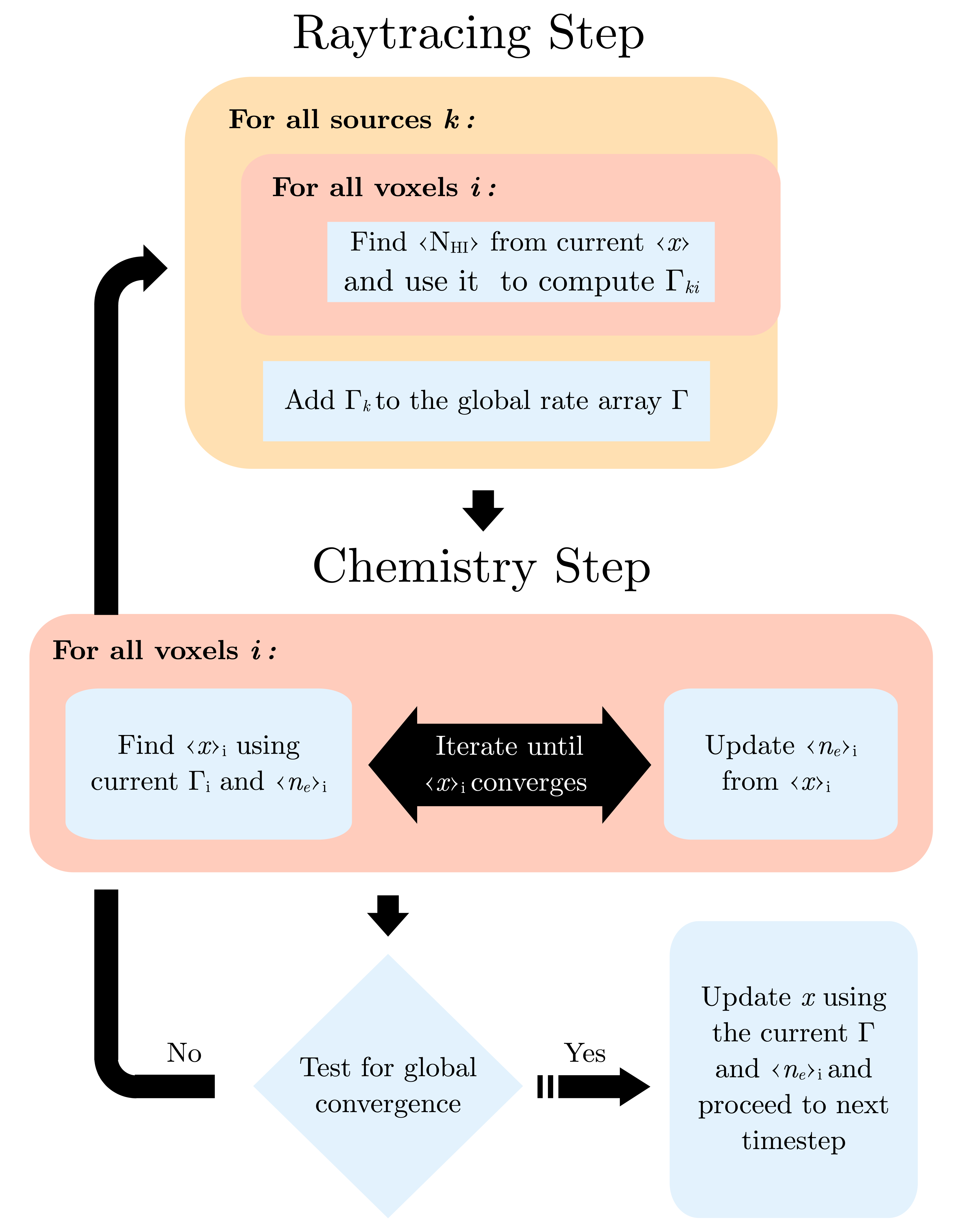}\vskip-2mm
    \caption{Flowchart representation of the method used by \ray{}. The figure shows the procedure for a single time-step in which the ionized fraction of hydrogen $x_{\rm HI}$ is evolved for the whole 3D grid. The method can be divided into a "ray-tracing" and a "chemistry" step, and multiple iterations of either typically occur in a single time step.}
    \label{fig:c2ray_flowchart}
\end{figure}

\ray{} overcomes this problem using an alternative approach, illustrated schematically in \autoref{fig:c2ray_flowchart}. As is argued in \citetalias{c2ray}, when recombinations and collisional ionizations are neglected, the solution of \autoref{eq:chemistry} over any timestep $\Delta t$ depends only on the time-averaged photoionization rate within that timestep, denoted by $\langle\Gamma\rangle$. Furthermore, only small deviations arise when collisions and recombinations are included, as is tested in \citetalias{c2ray}. The idea behind the \ray{} algorithm is to converge to the correct $\langle\Gamma\rangle$ within a given $\Delta t$ by iterating between a \emph{Ray-tracing Step}, which computes $\langle\Gamma\rangle$ based on the currently assumed solution for the time-averaged ionization state $\langle x_\mathrm{HII}\rangle$ of the whole medium, and a \emph{Chemistry Step}, which computes an updated $\langle x_\mathrm{HII}\rangle$ based on the new $\langle\Gamma\rangle$. This is illustrated in \autoref{fig:c2ray_flowchart} by the long vertical black arrow, which goes through a convergence test to determine whether the iteration needs to be repeated.

The chemistry step itself is not entirely trivial, as it still relies on being able to solve the differential equation. The method used in \ray{} is based on \citet{SchmidtVoigt1987}, who argue that when $n_e$, $\Gamma$, $C_{\rm H}$ and $\alpha_{\rm H}$ are assumed to be constant, an analytical solution exists for \autoref{eq:chemistry}. Using this solution, the time-averaged ionization state can be expressed as
\begin{align}
    \langle x \rangle &= x_{eq} + (x_0-x_{eq})(1-e^{-\Delta t/t_i})\frac{t_i}{\Delta t} \label{eq:time_avrg_sol} \\
    x_{eq} &= \frac{\Gamma + n_e\,C_{\rm H}}{\Gamma + n_e\left( C_{\rm H} + \alpha_{\rm H} \right)} \\
    t_i &= \left[\Gamma + n_e\left( C_{\rm H} + \alpha_{\rm H} \right) \right]^{-1}.
\end{align}
Here, $x_0$ is the ionization state at the beginning of the timestep, and $x_{eq}$ is the equilibrium solution, i.e., for $dx_\mathrm{HII}/dt = 0$, while $t_i$ is a constant time scale employed for the time-averaged inhomogeneous solution. Note also that $\Gamma$ has been used instead of $\langle\Gamma\rangle$ to represent the time-averaged photoionization rate to ease up notation. Since the non-time-averaged rate is never used in the algorithm, this new notation shall be used from now on. \ray{} uses this solution, and iterates for the electron density $n_e$ (which depends on $x_\mathrm{HII}$ through roughly $n_e \sim x_\mathrm{HII} n_\mathrm{H}$), until $\langle x_\mathrm{HII}\rangle$ converges. The thick horizontal arrow within the chemistry step illustrates this second iterative process in \autoref{fig:c2ray_flowchart}. Again, a convergence criterion is implicitly used to determine when to end the iteration.

The ray-tracing step requires further consideration, as it is the focus of the present work. It is discussed in detail in \S\ref{sec:computerate} and \S\ref{sec:ray-tracing_c2ray}. The main takeaway from this section is that the \ray{} method makes it possible to use very long timesteps while still remaining accurate even in the presence of fast-moving I-fronts.

\subsection{Computing Rates}\label{sec:computerate}
In the "ray-tracing step" introduced above, the properties of ionizing sources are used together with the knowledge of the ionization state of the medium inside a simulation volume to compute the photo-ionization rate $\Gamma$ occurring at any point in space. An ionizing source of specific luminosity $L_\nu$ located at a point $\vec{p}_\star$ produces at a target point $\vec{p}$ a (time-averaged) photo-ionization rate given by
\begin{equation}
    \Gamma = \frac{1}{4\pi r^2}\int_{\nu_{th}}^{\infty} \frac{L_{\nu}\sigma_\nu e^{-\langle\tau_\nu\rangle}}{h\nu}d\nu,
    \label{eq:gammalocal}
\end{equation}
where $\nu_{th}$ is the threshold frequency for photoionization ($h\nu_{th} = 13.6$ eV), $\left<\tau_{\nu}\right> \equiv \tau_{\nu}(r)$ is the time-averaged optical depth between source and target and $r = |\vec{p}_\star - \vec{p}|$ is the distance between $\vec{p}_\star$ and $\vec{p}$. The optical depth is proportional to the column density of neutral hydrogen $N_{\rm HI}$, and the proportionality factor is its frequency-dependent photoionization cross-section, $\tau_{\nu} = \sigma_{\nu} N_{\rm HI}$. The frequency-dependence of $\sigma_{\nu}$ approximately follows a power law whose index depends on the frequency band considered \citep[see, e.g.][for further details]{helium}.

\ray{} in its current form is a Cartesian grid code, which discretizes space using $N^3$ cubic \emph{voxels}, where $N$ is the number of voxels in each dimension. Using \autoref{eq:gammalocal} directly on a grid is problematic when the voxels are optically thick, i.e. the optical depth of a single voxel is non-negligible. The photoionization rate will then vary appreciably from one side of a voxel to the other. Using \autoref{eq:gammalocal} computed at an arbitrary point in the voxel as a representative rate for the whole voxel will, therefore, lead to an error in photon conservation - the number of ionizations occurring in the voxel will not be equal to the number of absorptions. To avoid this problem without being forced to use impractically small voxels, \ray{} works by \emph{imposing} that the number of ionizations is equal to the number of absorptions used to attenuate the radiation. As is detailed in \citetalias{c2ray}, using this condition leads to an alternative expression for the photoionization rate,
\begin{equation}
    \Gamma = \int_{\nu_{th}}^{\infty} \frac{L_{\nu}}{h\nu} \frac{e^{-\left<\tau_{\nu}\right>}(1-e^{-\Delta\tau_{\nu}})}{n_{\rm HI} V_{shell}}d\nu,
\label{eq:gamma}
\end{equation}
where $\Delta \tau_{\nu}$ is the optical depth \emph{through} the voxel, which is proportional to the light travel path length $ds$ through the voxel, and $\left<\tau_{\nu}\right>$ is the optical depth \emph{up to} the voxel. $n_{\rm HI}$ is the number density of neutral hydrogen inside the voxel, while the factor $V_{shell} = 4\pi r^2 ds$ accounts for both geometrical diffusion of radiation and the finite size of the cell. Note that, in the optically thin limit ($\Delta\tau_{\nu}\rightarrow 0$), the above expression reduces to \autoref{eq:gammalocal}. By defining the function
\begin{equation}\label{eq:tables}
    \gamma(N_{\rm HI}) \equiv
    \int_{\nu_{th}}^{\infty}
    \frac{L_{\nu} e^{-\sigma_{\nu} N_{\rm HI}}}{h\nu} d\nu,
\end{equation}
\autoref{eq:gamma} can be written in a more suggestive way
\begin{equation}\label{eq:sugg}
    \Gamma = \frac{1}{n_{\rm HI} V_{shell}}\left[\gamma(N_{\rm HI}) - \gamma(N_{\rm HI}+\Delta N_{\rm HI})\right].
\end{equation}
This means that rather than numerically solving the integral in \autoref{eq:gamma} each time it is required, the function $\gamma(N_{\rm HI})$ can be pre-calculated and tabulated for a range of column densities and a simple interpolation used to evaluate it for any given value of $N_{\rm HI}$. Note that the individual properties of the voxel where $\Gamma$ is computed are not part of the tabulation and are explicitly accounted for in \autoref{eq:sugg}.

When more than one source is present, the situation becomes slightly more complicated. The approach described in the original \ray{} paper (see Figure 4 in \citetalias{c2ray}) involves randomizing the order of sources and performing the chemistry step for each source individually before testing global convergence. However, this approach was modified in subsequent updates to the code, and it is this updated algorithm that we use here. The idea is simply to compute the 3D rate array (one rate per voxel) for each source and sum these arrays to obtain a global rate array. This global rate is then used for the chemistry step, and the process is repeated until convergence. Note that this is the process as illustrated in \autoref{fig:c2ray_flowchart}, where we use the notation $\Gamma_{ki}$ to signify that this quantity applies to a given source indexed by $k$ and a given voxel, indexed by $i$.

\section{Novel ray-tracing Method: \texttt{ASORA}} \label{sec:ray-tracing}
As shown by \autoref{eq:gamma}, the problem of finding ionization rates boils down to computing the column density $N_{\rm HI}$ of neutral hydrogen between a source and grid voxels. This is the process we refer to as \emph{ray-tracing} in this context. In principle, given a cubic grid with $N$ voxels in each dimension, it is possible to compute $N_{\rm HI}$ directly for all voxels of the grid, an approach known as ``long characteristics'' (LC). For a single source, it scales as $\mathcal{O}(N^4)$. This is because, for each of the $N^3$ voxels to treat, the number of other voxels that lie along the ray coming from the source is on the order $\mathcal{O}(N)$. LC has the advantage of being easy to parallelize as all rays are treated independently. However, given that radiation propagates causally outward from the source and that column density is an additive quantity along a given line of sight, this algorithm also contains a lot of redundancy. A variety of methods have been proposed to make ray-tracing more efficient \citep[see, e.g.][for an overview]{Ramses-RT}. \ray{} uses a version of the ``short-characteristics'' (SC) ray-tracing method \citep{raga1999}, which reduces the redundancy of the problem by using interpolation from inner-lying voxels relative to the source to compute the column density to outer-lying ones. This method reduces the complexity to $\mathcal{O}(N^3)$ but is harder to parallelize as it introduces voxel dependency.

Since the effect of each source is independent, the total cost of the ray-tracing step is the number of sources $N_\mathrm{src}$ times whatever the cost for a single source is, e.g., $\mathcal{O}(N^4)$ for LC or $\mathcal{O}(N^3)$ for SC. On the other hand, the total cost of a chemistry step only scales with the number of voxels in the grid, i.e., $N^3$. This clarifies why the ray-tracing step is the primary target for optimization in an EoR code like \ray{}, where typically $N_\mathrm{src} \gg 1$. In fact, at low redshift, it is common to have a source in almost every voxel, so that $N_\mathrm{src}\sim N^3$, and that, in turn, the complexity of the ray-tracing step is $\sim\mathcal{O}(N^6)$ (using SC) versus $\mathcal{O}(N^3)$ for the chemistry step.

We should also mention that we never separately treat more than one source per voxel and instead simply add the luminosity of all sources whenever more than one is present, implying $N_\mathrm{src} \leq N^3$. The result is identical to treating them separately and adding up the resulting rates because a source is always assumed to be at the center of a voxel. Note, however, that this approach is possible only because the spectra of all sources are identical, and in the future, when different source types may be considered by the code, this procedure will have to be adapted by only summing up the sources that belong to the same type.

Below, we first discuss in detail the short-characteristics ray-tracing method used in \ray{} (\S~\ref{sec:ray-tracing_c2ray}). Then, we give an overview of the CPU-parallelization strategy the code has used so far (\S~\ref{sec:CPU_parallelization}). Next, we introduce the adaptation of the method for GPUs (\S~\ref{sec:GPU_parallelization}), and finally, in \S~\ref{sec:pyton_wrapping}, we discuss the structure of the new \texttt{Python} wrapper built around \ray{}.

\subsection{Ray-tracing in C$^2$Ray}\label{sec:ray-tracing_c2ray}
Here, we closely follow the discussion of Appendix A in \citetalias{c2ray}, and in particular, refer the reader to Figure A1, which provides a good visual description of the geometric arguments detailed below. For a voxel located at mesh position $p=(i,j,k)$ and a source at $s=(i_s,j_s,k_s)$, the full column density $N_{\rm HI}=\tilde{N}_{\rm HI} + \Delta N_{\rm HI}$ along the ray from $s$ to $p$ can be decomposed into a part \emph{up to} the voxel $\tilde{N}_{\rm HI}$ and a part \emph{within} the voxel $\Delta N_{\rm HI}$. The latter is proportional to the physical path length $dl = \alpha\,ds$ through the voxel at $p$, where $ds$ is the path length in mesh units and $\alpha$ is the physical length of a grid voxel,
\begin{equation}
    \Delta N_{\rm HI} = n_{\rm HI}\,\alpha\,ds.
\end{equation}
Defining $\Delta i = i - i_s$ (and similarly $\Delta j$ and $\Delta k$), one can determine through which plane the ray coming from $s$ enters the voxel at $p$. For example, if $\Delta k > \Delta i$ and $\Delta k > \Delta j$, the ray enters through one of the constant-$z$ planes, with the $\Delta k$ sign indicating which one. In this particular case, the path length through the voxel is
\begin{equation}
    ds = \sqrt{1+\frac{\Delta i^2 + \Delta j^2}{\Delta k^2}},
\end{equation}
and the analogous expressions apply if the ray enters through the constant-$x$ or $y$ plane. The main assumption of the short-characteristics method is that $\tilde{N}_{\rm HI}$ can be computed by interpolation with neighboring voxels of $p$ that are closer to $s$. The particular scheme used by \ray{} \citep{raga1999} uses 4 neighbours, whose positions are given by

\begin{figure}[t]
    \centering
    \includegraphics[width=\linewidth]{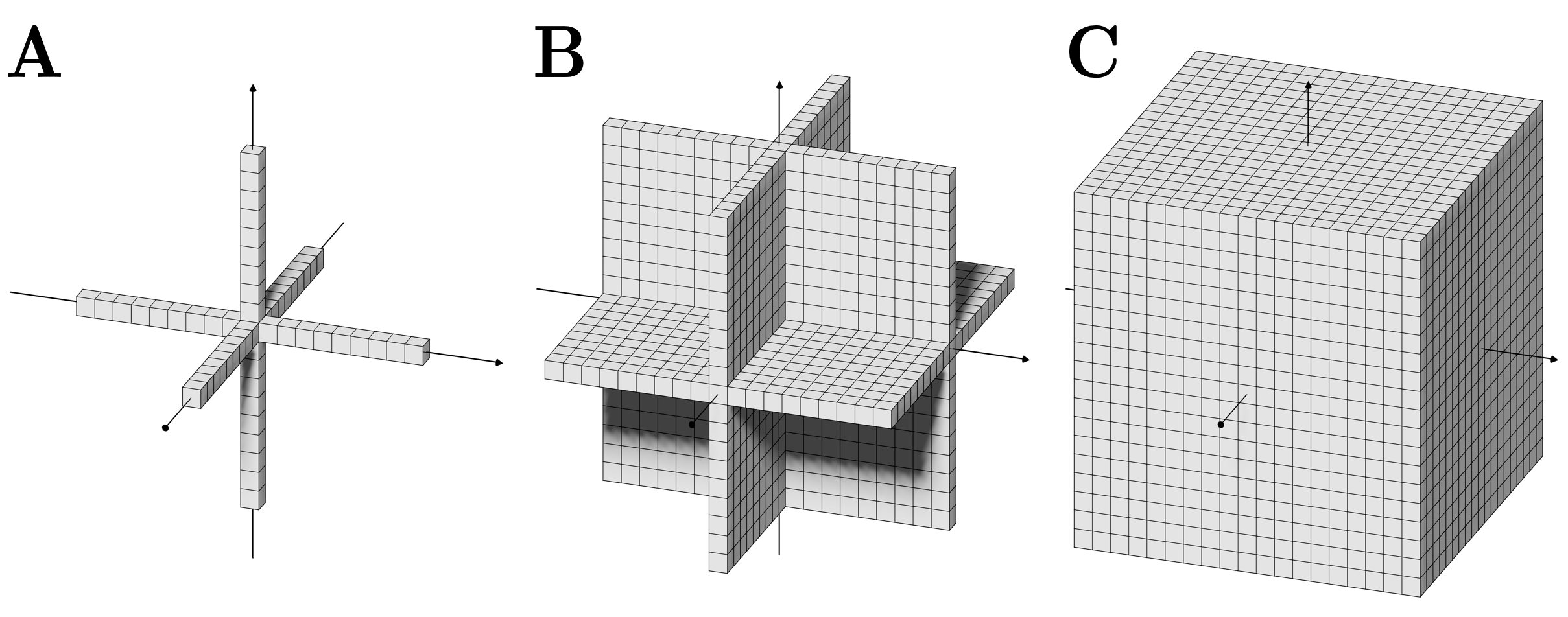}
    \caption{Parallelization strategy used by the original \ray{} code. In the first step (A), 6 grid domains can be treated independently, corresponding to axes around the source voxel. In (B), the 12 planes joining them form independent domains, while in the third one (C), the 8 octants between the planes do.}
    \label{fig:c2ray_dd}
\end{figure}

\begin{align}
    \begin{split}
        e_1 &= (i,j,k-\sigma_k), \quad \quad
        e_2 = (i,j-\sigma_j,k), \\
        e_3 &= (i-\sigma_i,j,k), \quad \quad
        e_4 = (i-\sigma_i,j-\sigma_j,k-\sigma_k) \,
    \end{split}
    \label{eq:raga}
\end{align}
where $\sigma_{i,j,k} = \frac{|\Delta i,j,k|}{\Delta i,j,k}$. The interpolated column density up to $p$ then reads
\begin{equation}\label{eq:interp}
    \tilde{N}_{\rm HI} = w_1N_{e_1} + w_2N_{e_2} + w_3N_{e_3} + w_4N_{e_4}.
\end{equation}
The interpolation weights $w_n$ are a simple geometric weighting based on the $xy$-distance from the corner to the point of intersection between the ray and the surface of the cell. They are chosen such that when the ray is parallel to an axis or lies on a grid diagonal, in which case $\tilde{N}_{\rm HI}$ is exactly equal to the column density of only one of the neighbors, all but the weight of that neighbor vanish. The interested reader is referred to Appendix A in \citetalias{c2ray} for the details of this weighting choice.

The above scheme describes how the column density up to a given voxel can be approximated using the knowledge of the equivalent quantity corresponding to 4 other voxels that lie closer to the source. This inter-voxel dependency naturally implies that for the scheme to be applied correctly, one must treat the voxels in a particular order, starting at the source voxel and moving outward from there. This ensures that the interpolation step does not attempt to use information that doesn't exist yet, so we say that SC is a \emph{causal} algorithm. In fact, the simplest way to traverse the grid is to simply perform a triple loop over the $x\rightarrow y\rightarrow z$ indices of all voxels by starting the loop at the source voxel indices. This is a fully sequential approach. The next two sections deal with the problem of finding parallel alternatives to the latter.

\subsection{Existing CPU Parallelization and Optimizations}\label{sec:CPU_parallelization}
The current version of \ray{} uses various methods to optimize the cost of ray-tracing and make the procedure scalable to massively parallel CPU systems. A key feature of the code is that the treatment of each source is completely independent. \ray{} harnesses this independence by distributing the full list of sources between MPI ranks. Each rank receives a copy of the full grid data and works on a subset of the sources. It performs ray-tracing for each source in this subset and sums together their respective ionization rate arrays (see \S\ref{sec:computerate}). Then, an MPI reduction operation is used to sum the rate arrays of all ranks and obtain a global $\Gamma$ that includes the contributions of all sources. This allows the full ray-tracing workload to be distributed over many processors in shared and distributed memory setups. The main limitation of this setup is memory since each rank carries a full copy of the 3D grid.
\begin{figure}[t]
    \includegraphics[width=\linewidth]{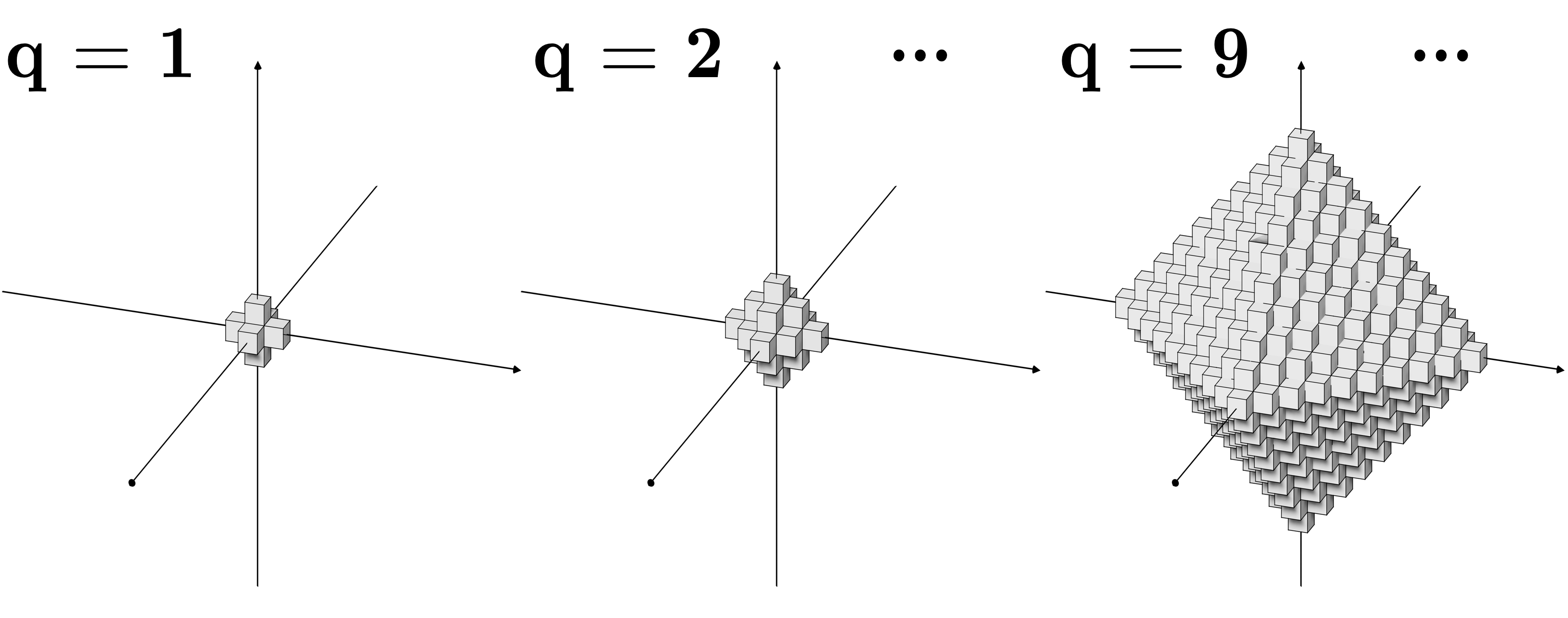}\vskip-4mm
    \caption{Sequence of octahedral shells $S_q$ used in the \texttt{ASORA} ray-tracing method. All voxels belonging to a shell $S_q$, with $q>0$, depend strictly on voxels from previous shells $\left\{S_{r}\,|\, r < q\right\}$. The shells $q=1$, $q=2$ and $q=9$ are shown, with the source voxel ($q=0$) at the origin of the axes.}
    \label{fig:octa}
\end{figure}
As was explained in \S\ref{sec:ray-tracing_c2ray}, the ray-tracing work for a single source is more challenging to do in parallel due to the inter-voxel dependency of the SC method. However, it is possible to find independent subdomains of the grid and use an approach similar to domain decomposition. This approach performs the following steps in order, which are illustrated in \autoref{fig:c2ray_dd}:
\begin{enumerate}
    \item Do the 6 axes outward from the source voxel (\textbf{A}) in parallel
    \item Do the 12 planes joining these axes (\textbf{B}) in parallel
    \item Do the 8 octants between the planes (\textbf{C}) in $x\rightarrow y\rightarrow z$ order, in parallel,
\end{enumerate}
where the labels \textbf{A}, \textbf{B} and \textbf{C} correspond to the three sketches in \autoref{fig:c2ray_dd}. \ray{} uses OpenMP tasks to do the independent domains following this approach, which can yield a speedup of $S \lesssim 8$.
\begin{figure*}[h!]
    \centering
    \includegraphics[width=0.85\linewidth]{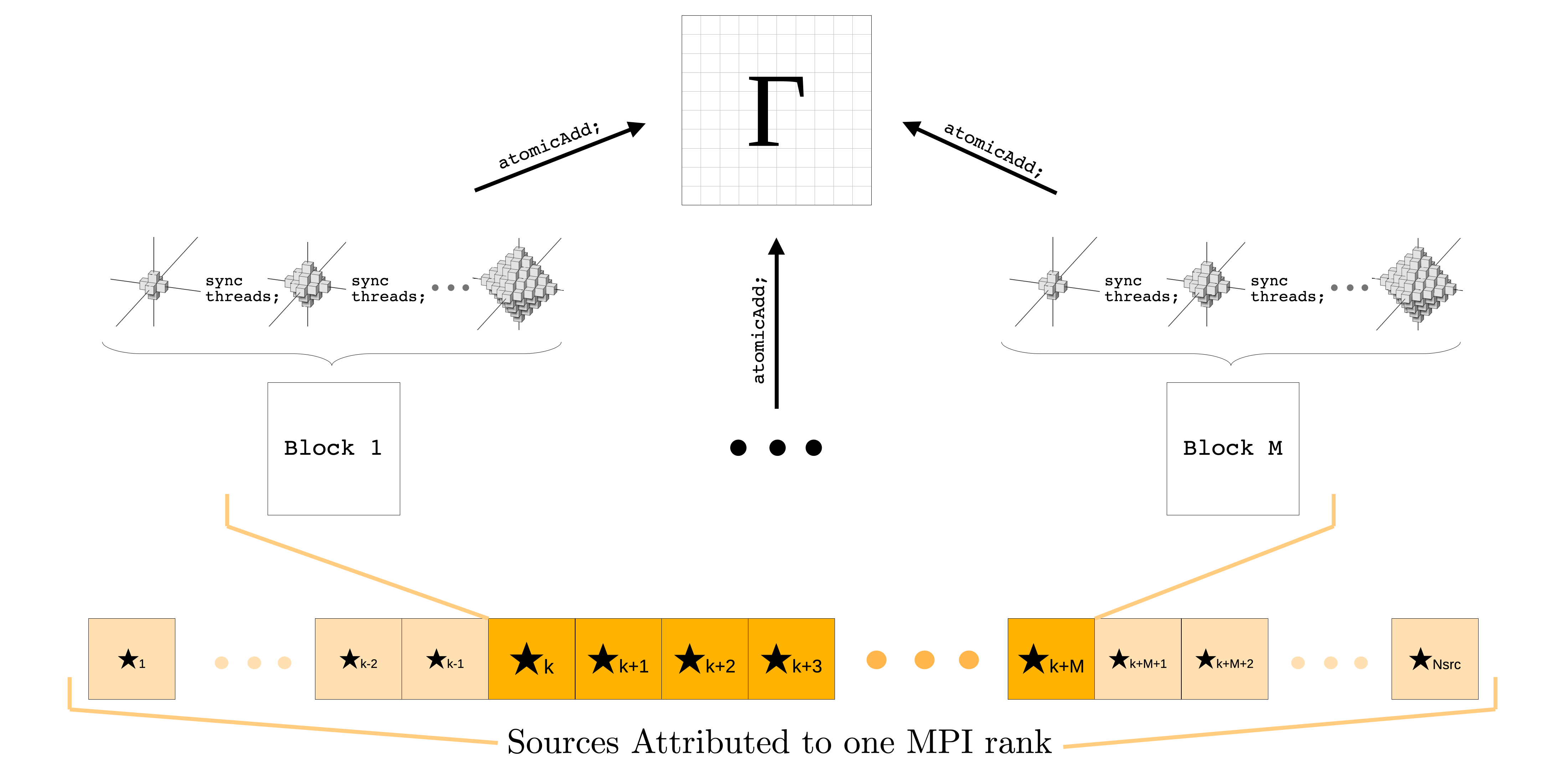}
    \caption{Implementation of the \texttt{ASORA} method. $N_\mathrm{src}$ sources, labeled by $\star_i$, are treated in batches of a given size $M$, and one block is dispatched for each source in the batch. Threads within a block are synchronized between each shell $S_q$ (see \autoref{fig:octa}) but are independent across different sources. Each block atomically contributes to the global ionization rate array. In MPI mode, each rank independently follows the above framework and the $\Gamma$ arrays of all ranks are sum-reduced to the root.}
    \label{fig:blocks}
\end{figure*}
Finally, the ray-tracing procedure itself is optimized using the following technique: rather than ray-tracing the whole grid, the program first treats only a cubic sub-region around the source, namely a "sub-box", and then calculates the total amount of radiation that leaves this sub-box (i.e. a photon loss). If this loss is above a given threshold, the program increases the size of the sub-box, treats the additional voxels and repeats this procedure until the photon loss is low enough. This allows \ray{} to avoid expensively ray-tracing all voxels when, in fact, almost no radiation reaches the ones far away from the source. The threshold value should be chosen based on convergence studies of the type of problem being simulated. The sub-box technique has been found to work well in EoR settings, where the density field is almost Gaussian. In some more specific situations, where narrow, optically thin tunnels exist in otherwise optically thick regions, the technique might produce inaccurate results. In these cases, using a very small threshold value or, in the worst case, ray-tracing the whole grid may be desirable. Additionally, the user can impose a hard limit on the maximum distance any photon can reach relative to the source.

\subsection{GPU Implementation}\label{sec:GPU_parallelization}
GPUs are designed to execute numerous concurrent operations, organized into units referred to as \textit{blocks} in CUDA and \textit{workgroups} in AMD terminology. Given that \texttt{ASORA} has been implemented using CUDA, we will continue to use CUDA terminology. We are also planning a future port of the library for AMD platforms. Threads can be synchronized within a block, while blocks run asynchronously \citep{Nickolls2008Scalable}. It is possible to perform a synchronization between blocks only globally. To fully harness the resources of a GPU, one aims to ensure that the number of threads active at any given time is as close as possible to the theoretical maximum of the used device. While no universal prescription exists to achieve this, it is generally desirable that blocks have a similar workload and their number is in the same order as the number of streaming multiprocessors (SMs) available on the device. This suggests a natural implementation for the ray-tracing problem: dispatch one block for each source and use intra-block synchronization to respect the causality of the short characteristics algorithm. 

For this approach to be efficient, however, the work for a single source cannot be simply parallelized following the domain decomposition approach described in \S~\ref{sec:CPU_parallelization} as this would allow at most 8 threads to be active within a block. To parallelize the work for a single source in a way more suited for the capabilities of a GPU, we recall that radiation would propagate as a spherical wavefront around a point source in a continuous medium. This translates to a series of shells around a source voxel in the discretized setting. It turns out that there is a particular sequence of disjoint shells $S_q$, illustrated in \autoref{fig:octa}, which are causally ordered with respect to the SC scheme used by \ray{}. $q$ indexes the "distance" of the shell to the source; the $q=0$ shell is simply the source voxel itself, and $q=1$ contains the 6 directly adjacent voxels to the source. The causal ordering can be summarized by the following conditions:
\begin{enumerate}
    \item The first shell $q=0$ contains only the source voxel, which can be treated directly without interpolation.
    \item For any voxel $p \in S_q$ with $q>0$, all 4 interpolation neighbors appearing in \autoref{eq:raga} belong strictly to shells $S_{r}$ with $r<q$, in other words, only to shells "below" the current one.
    \item In particular, all voxels $p \in S_q$ are independent of one another with respect to the interpolation scheme.
\end{enumerate}

This means that the full ray-tracing work for a single source can be divided into the sequence of tasks $\{S_q\}_{q=0}^Q$, where $Q$ is the size of the largest shell. These $Q$ tasks must be done sequentially by definition, but each task comprises subtasks (one subtask for each voxel in the shell) that can be performed in parallel. Note that the number of voxels inside a shell $S_q$, and hence the number of independent subtasks per task, is $n_q = 4q^2 + 2$. Going back to the discussion above, when, for instance, 100 threads are assigned per source for each task with $q\geq 5$, it is theoretically possible for all threads to be actively engaged in performing work. This effectively resolves the challenge of parallelizing the computation on a per-source basis. 

Rather than giving a maximal shell size $Q$, it is more convenient to set a maximum physical radius $R_{\gamma}$ any photon can travel from the source. If the physical size of a grid voxel is, as previously, denoted by $\alpha$, the (dimensionless) size index of the largest shell required to cover the chosen radius fully is given by $Q = \lceil\frac{R_{\gamma}}{\alpha \sqrt{3}}\rceil$. Any cell inside $S_Q$ whose distance to the source exceeds $R_{\gamma}$ can simply be excluded from the computation to yield a spherical region in which $\Gamma$ is nonzero.

The full implementation, illustrated in \autoref{fig:blocks}, goes as follows. We dispatch one CUDA thread block for each source that works through the sequence of shells. The result, i.e. the photo-ionization rate produced by that source in each voxel, is \emph{atomically} added to the global rate array $\Gamma$. In practice, there is a small additional caveat to consider, namely that by the nature of the algorithm, each source requires a temporary memory space to store the values of the previously interpolated voxels needed for the next interpolation. The required space can typically be a good fraction of the whole grid, so the number of blocks that can be dispatched together is limited by GPU memory. In fact, rather than directly dispatching one block for each of the $N_\mathrm{src}$ sources in the simulation, we group the sources into batches of size $M$, and work on these batches one after the other. $M$ is determined by available GPU memory and grid size $N$. As long as $M$ is large enough to saturate the GPU, this approach should not result in a significant performance loss compared to the ideal scenario of immediately deploying one block per source, without any batching, since the workload for each source is the same.

Finally, \texttt{ASORA} is also MPI-enabled, using \texttt{mpi4py} \citep{mpi4py} in the same way as it was intended for the original \ray{}. Namely, the sources are evenly distributed to multiple MPI processes. Each MPI rank maps to one GPU, which then uses the model laid out above to process its subset of sources and broadcasts the result $\Gamma$ to the root rank, using \texttt{MPI\_REDUCE} with a sum operation. This allows using \texttt{ASORA} on a multi-GPU setup across multiple nodes to further speed up ray-tracing on very heavy workloads.

To conclude this section, we note that the \texttt{ASORA} method, as presented here, only applies to uniform grids, as the octahedral shell approach builds on this assumption. We acknowledge that this is a strong limitation of our method, and we plan to explore its adaptability to non-uniform grids. Further technical details on \texttt{ASORA} can be found in \ref{sec:cuda}.

\subsection{Python-wrapping of \ray{}}\label{sec:pyton_wrapping}
\begin{figure}[h]
\centering
    \includegraphics[width=0.95\linewidth]{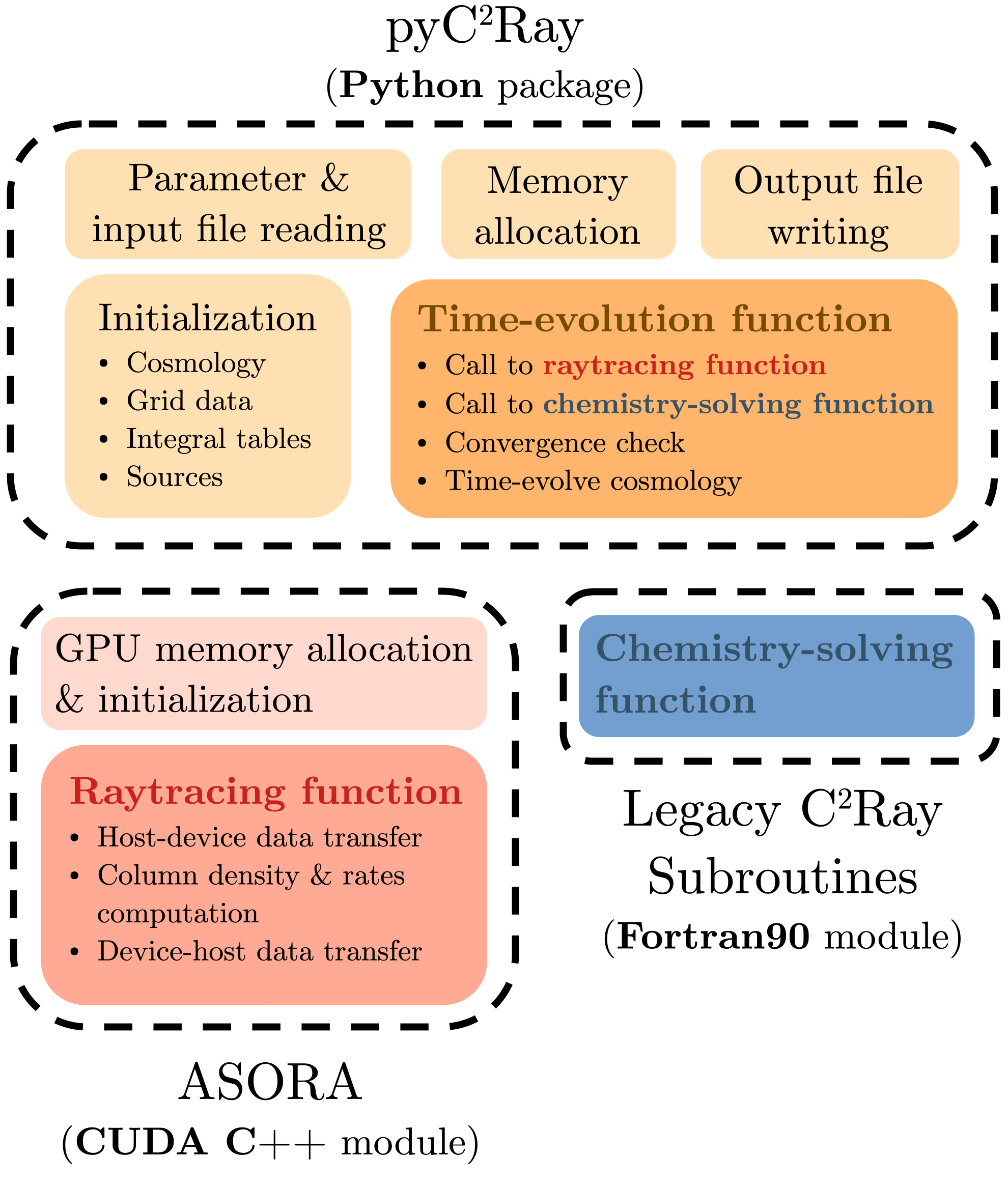}
    \caption{Structure of the \pyray{} code. The main python package, \texttt{pyc2ray}, sets up the simulation and acts as the front end to the user. Internally, the time-evolution method of this package executes functions from two compiled extension modules. One is \texttt{ASORA}, the new GPU ray-tracing module written in CUDA C++, while the other contains a set of wrapped \texttt{Fortran} subroutines taken and adapted from the original \ray{} code.}
    \label{fig:pythonwrapping}
\end{figure}
Here, we provide a brief overview of the \pyray{} interface and architecture, summarized visually in \autoref{fig:pythonwrapping}. The package amalgamates key components from the original \texttt{Fortran90} code, the new ray-tracing library as discussed above, and elements of pure \texttt{Python}. This integration is facilitated through \texttt{f2py}, a tool developed as part of the \texttt{NumPy} project \citep{numpy2020}. This tool streamlines the creation of extension modules from \texttt{Fortran90} source files.

The incorporated \texttt{Fortran90} subroutines primarily encompass the chemistry solver and retain the original CPU-based ray-tracing module as a contingency. The novel \texttt{ASORA} method is written in \texttt{C++/CUDA} and compiled as a \texttt{Python} extension module natively compatible with \texttt{NumPy}. The principal time-evolution function within \pyray{} is implemented in \texttt{Python}, and it invokes the ray-tracing method, choosing between the CPU and GPU versions and the chemistry method sourced from these extension modules. The prior process of precalculating photoionization rate tables, as introduced earlier, has transitioned to direct implementation in \texttt{Python}. This is achieved using numerical integration techniques from the \texttt{SciPy} library \citep{scipy2020}, which relies on the underlying \texttt{QUADPACK} library for lower-level computations. It is worth noting that these integration methods differ from the custom Romberg integration subroutines utilized by the original \ray{} framework. The commonly needed cosmological equations and physical quantities are now provided by \texttt{Astropy} \citep{astropy2022}.

Beyond these technical aspects, the inherent method within \pyray{} —apart from the ray-tracing component— has undergone minimal alteration. Key features of \ray{}, including photoionization and hydrogen chemistry, have been seamlessly migrated to the \texttt{Python} version without compromising computational efficiency. Our strategy involves a gradual integration of additional extensions over time.

\section{Validation Testing \& Benchmarking}\label{sec:results}
In \S~\ref{sec:accuracy_tests}, we validate our new code using a series of well-established tests, comparing our results to analytical solutions and to the results of our original \ray{} code. In \S~\ref{sec:performance_benchmark}, we investigate how the updated ray-tracing method scales relative to the main problem parameters. In all tests, the temperature conditions of the gas are assumed to be isothermal, i.e., no heating effects are modeled.

\subsection{Accuracy Tests}\label{sec:accuracy_tests}
We begin by conducting Tests 1 and 4 from \citetalias{c2ray}, labeled as \textit{Test 1} and \textit{2} here, to evaluate the precision of our code in monitoring I-fronts in single-source mode. This evaluation encompasses scenarios both with and without cosmological background expansion. Following this, we investigate the interplay among multiple sources and the occurrence of shadow formation behind an opaque object, \textit{Test 3} and \textit{Test 4}.
\begin{figure}[t]
    \centering
    \includegraphics[width=\linewidth]{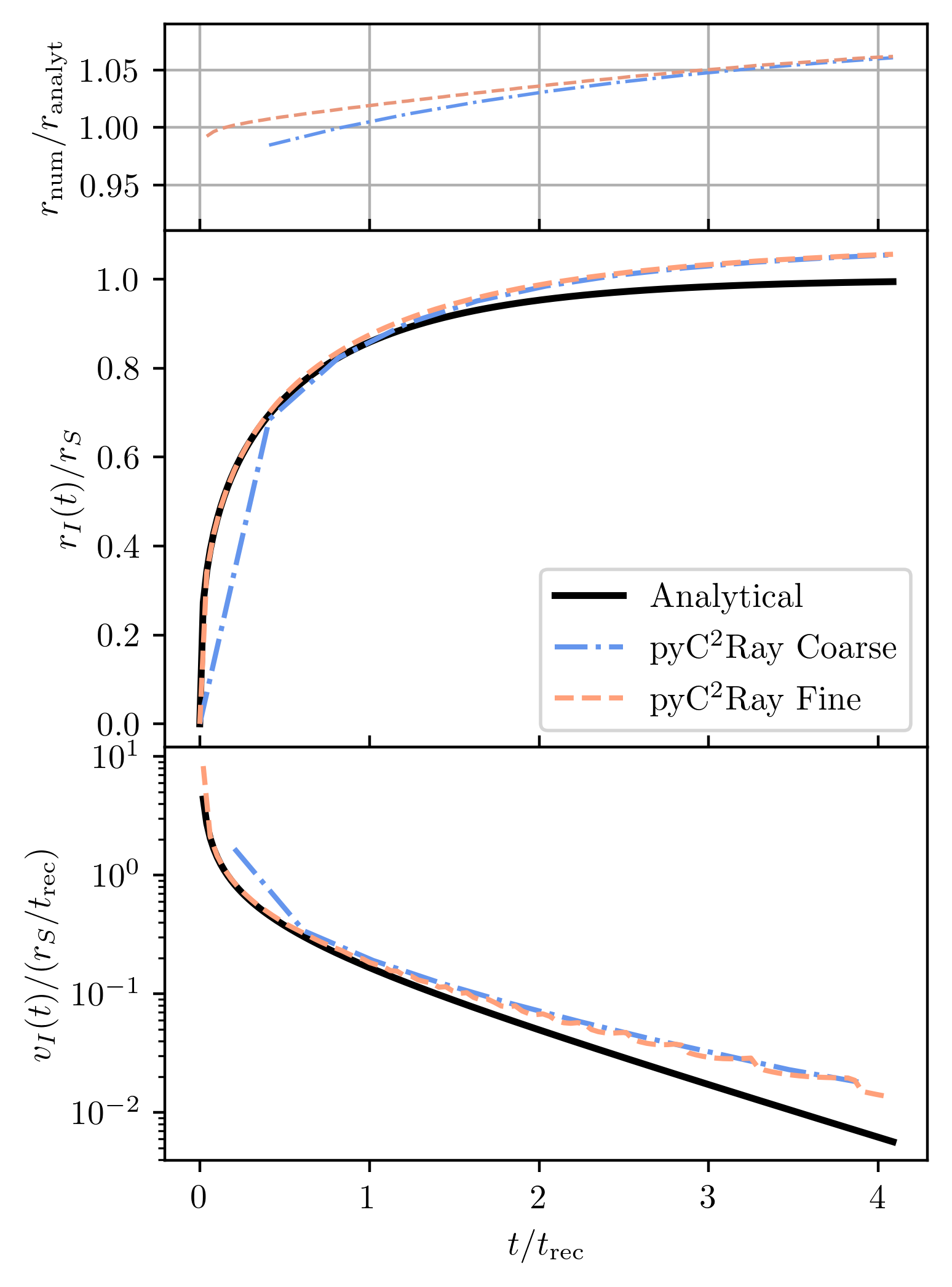}\vskip-4mm
    \caption{Result for Test 1 (Single-source H II region expansion in uniform gas). The test is conducted with a "coarse" time step $\Delta t_{c} = t_{evo}/10$ and a "fine" one, $\Delta t_f = t_{evo}/100$. The time evolution of the ionization front radius (\emph{middle}) and velocity (\emph{bottom}) are shown. The error between the numerical and analytical results can be seen in the top panel.}\label{fig:test1}
\end{figure}
\subsubsection{Test 1: Single-Source \hii{} Region Expansion} \label{sec:test1}

\begin{figure}
    \centering
    \includegraphics[width=\linewidth]{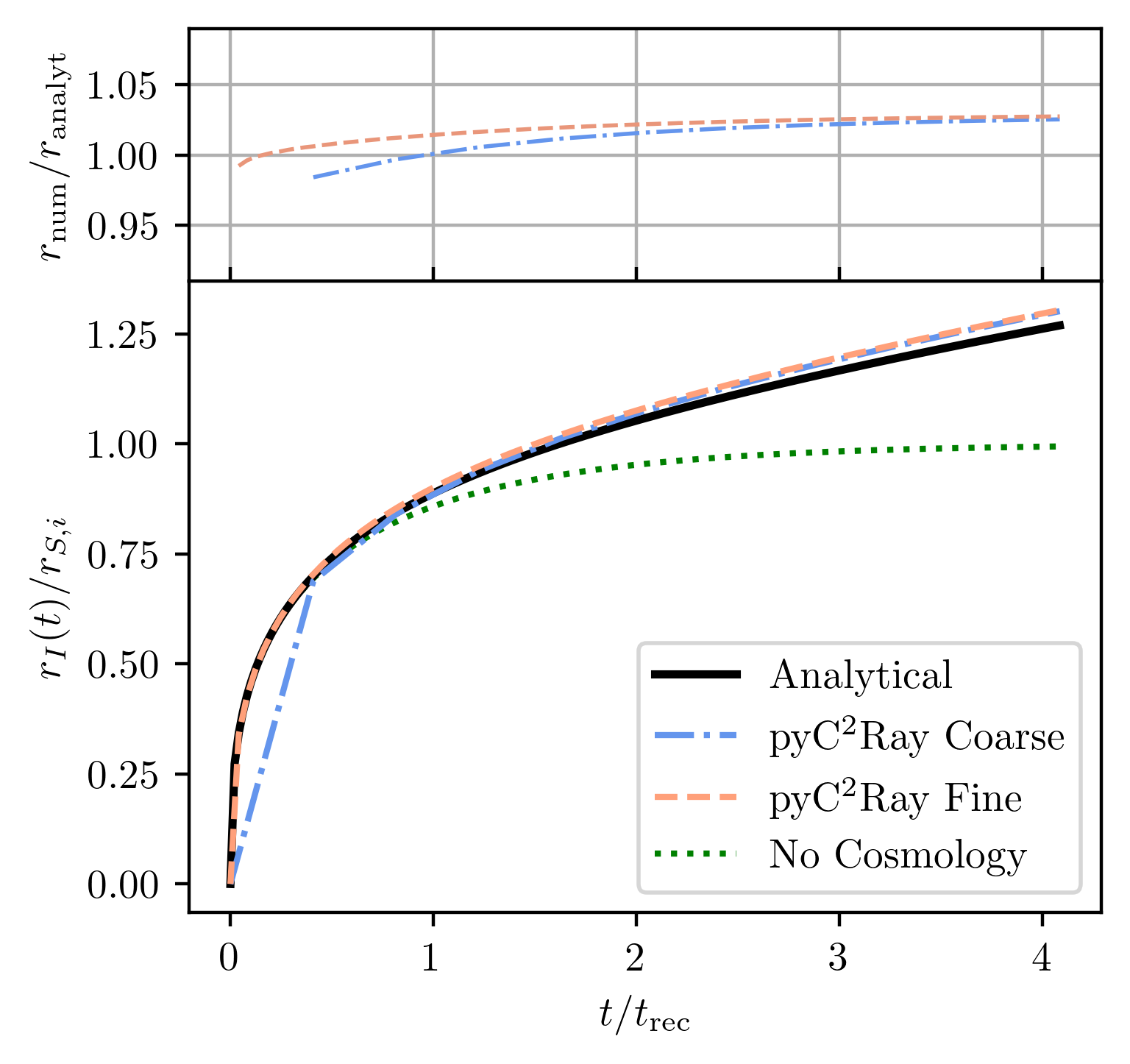}\vskip-4mm
    \caption{Result for Test 2 (Single-source \hii{} region expansion in cosmological expanding background). Notation is the same as in \autoref{fig:test1}. The source turns on at $z_i=9$, and the I-front radius is given in comoving kpc, with the scale factor $a(t_i)=1$, normalized to the instantaneous Strömgren radius at $z_i$, $r_{S,i}$}. The green dotted line shows the analytical result without cosmological expansion for reference.
    \label{fig:test4}
\end{figure}

Consider the classical scenario of a single ionizing source within an initially-neutral, uniformly dense field at a constant temperature. In this case, any cosmological effects are disregarded. Assuming the photoionization cross section remains frequency-independent, $\sigma_{\nu} = \sigma_{0}$, known as \textit{grey opacity}, this system has a well-established analytical solution for the velocity and radius of the ensuing ionization front with respect to time. The solution is given by
\begin{align}
    \label{eq:rI1}
    r_\mathrm{I}(t) &= r_\mathrm{S}\left[1-\exp(-t/t_{\rm rec})\right]^{1/3} \\
    v_\mathrm{I}(t) &= \frac{r_\mathrm{S}}{3\,t_{\rm rec}}\frac{\exp(-t/t_{\rm rec})}{\left[1-\exp(-t/t_{\rm rec})\right]^{2/3}} \ .
\end{align}
The above expressions depend on the Strömgren sphere radius $r_\mathrm{S}$, recombination time $t_\mathrm{rec}$ and luminosity emitted by the source (or the number of photons per unit time). These quantities are defined as,
\begin{align}
    r_\mathrm{S} =& \left(\frac{3\,\dot{N}_{\gamma}}{4\,\pi\,\alpha_\mathrm{H}(T)\,n_\mathrm{H}^2}\right)^{1/3} \ , \\
    t_{\rm rec} =& \frac{1}{\alpha_\mathrm{H}(T)\,n_{\rm H}} \ , \\
    \dot{N}_{\gamma} =&
    \int_{\nu_{\rm th}}^{\infty}
    \frac{L_{\nu}}{h\nu} d\nu \ .
    \label{eq:ngammalnu}
\end{align}
\begin{figure*}[h]
    \centering
    \includegraphics[width=\linewidth]{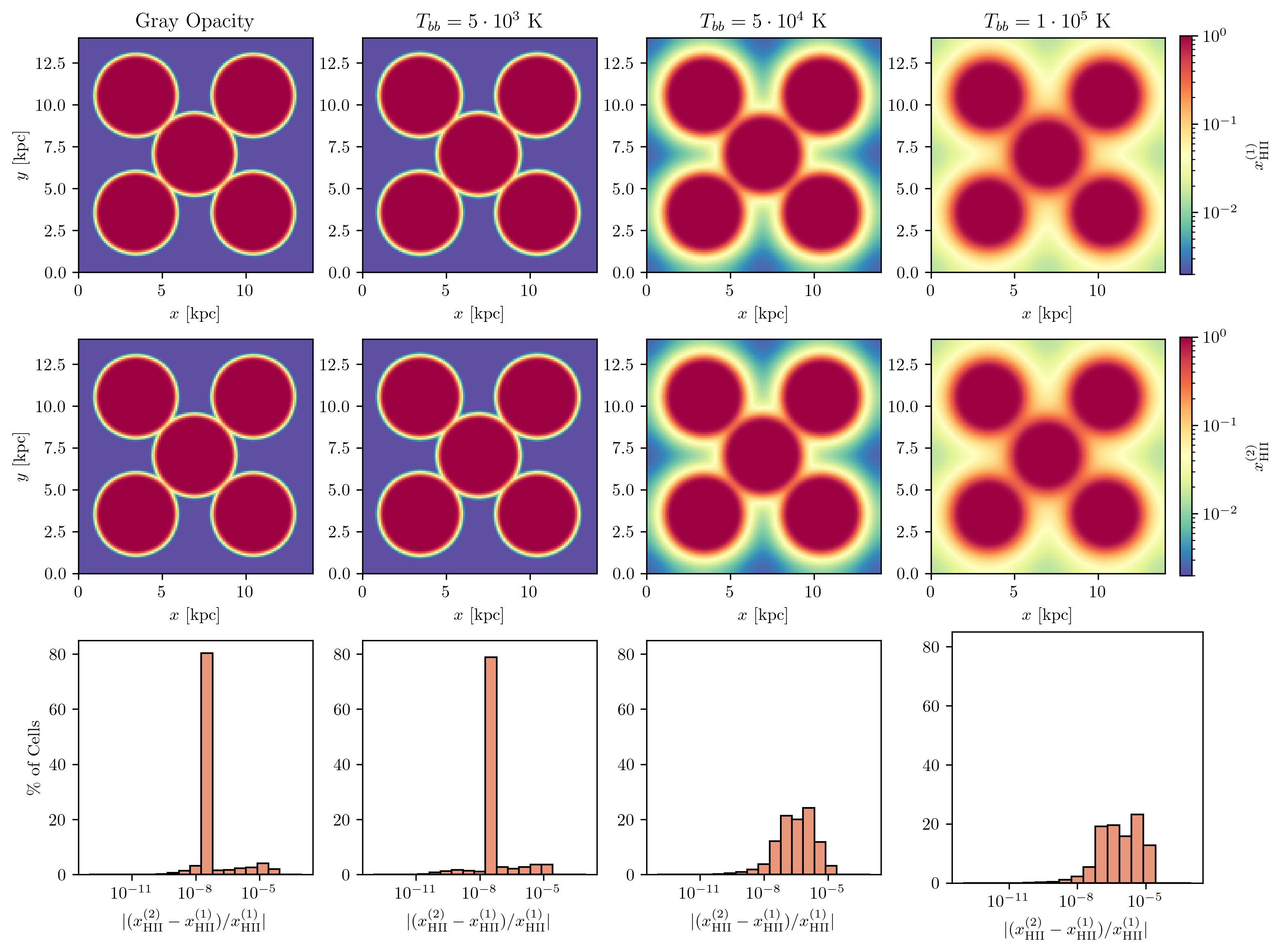}\vskip-4mm
    \caption{Result for Test 3 (Expansion of Overlapping H II regions around Multiple Black-Body Sources). The top and middle rows show slices through the simulation domain at the $z$-coordinate of the 5 sources, for \ray{} and \pyray{} respectively. The leftmost column corresponds to the case with grey opacity, and the remaining 3 columns to those where black body spectra with different temperatures $T_{bb}$ were used. Colors are normalized across each row. The bottom row shows the distribution of relative per-voxel errors between the 2 codes for the whole 3D grid in all 4 cases.}
    \label{fig:multisource}
\end{figure*}
Here, $L_\nu$ is again the specific luminosity of the source (power per unit frequency), which is related to the luminosity $\dot{N}_{\gamma}$ (number of ionizing photons per unit time) through \autoref{eq:ngammalnu}. We conduct our first test using the following numerical parameters: the luminosity of the source is $\dot{N}_{\gamma} = 10^{48}$ s$^{-1}$, the number density of hydrogen $n_\mathrm{H} = 10^{-3}$ cm$^{-3}$, its temperature $T=10^4$ K and the simulation box size is $10\,{\rm kpc}$. As stated above, we use the case B recombination coefficient for Hydrogen, $\alpha_\mathrm{H}(T=10^4\, {\rm K}) = 2.59\times 10^{-13}\,\rm cm^3\,s^{-1}$. Using these parameters, the recombination time is $t_\mathrm{rec} \simeq 122.35$ Myr and the Strömgren radius is $r_S = 3.15\, {\rm kpc}$. The simulation is run with mesh size $256^3$ for $t_{\rm evo} = 500\,\mathrm{Myr} \approx 4t_\mathrm{rec}$, following the prescription of Test 1 in \citet{comparison_1}. As in \citetalias{c2ray}, the simulation is repeated once with a coarse time step $\Delta_t = 50\,\rm Myr$ and once with a fine one, $\Delta t = 5\,\rm Myr$. We track the position of the I-front along the $x$-axis and define $r_\mathrm{I}(t_k)$ as the radius where $x_{\rm HI}=0.5$. The precise location within a voxel is found by linear interpolation. The numerical I-front velocity, $v_\mathrm{I}$, is found by finite-differencing $r_\mathrm{I}$, using the same approach as in \citetalias{c2ray}.

The results are shown in \autoref{fig:test1}, where the three panels contain the time evolution of the ratio between numerical to analytical results (top), the I-front radius (middle) and its velocity (bottom).
At times $t\lesssim t_\mathrm{rec}$, \pyray{} is in excellent agreement with the analytical prediction, both with a coarse and a fine time step choice. At $t\gtrsim t_\mathrm{rec}$, the numerical I-front overestimates the analytical prediction by as much as 6\%. This is consistent with the findings of, e.g., \citet{comparison_1}, where all tested codes predict such an overestimate. \citet{PawlikShaye2008} have demonstrated that this is because, in reality, the ionized fraction varies smoothly within the ionized bubble, whereas the Strömgren argument assumes a sharp transition from fully ionized to fully neutral.

\subsubsection{Test 2: Single-Source \hii{} Region in expanding background} \label{sec:test2}
We next test if \pyray{} correctly models the propagation of I-fronts in an expanding universe. Test 2 uses the same source parameters as Test 1, with the source turning on at $z=9$ and then shining for 500 Myr, while the background density starts with the same value as before and evolves with the expansion of the universe.
\citet{cosmicstromgren} showed that a generalized analytical solution exists in this case. The comoving I-front radius is given by $r_\mathrm{I}(t) = r_{\mathrm{S},i}\,y(t)^{1/3}$, where $r_{\mathrm{S},i} = (3\dot{N}_{\gamma}/4\pi\alpha_\mathrm{H}(T)n_{\mathrm{H},i}^2)^{1/3}$ is the instantaneous Strömgren radius at the ignition time, $t_i$, of the source (with the scale factor set to unity at $t_i$, $a_i=1$), and
\begin{equation}
    y(t) = \lambda e^{\lambda t_i / t}\left[\frac{t}{t_i} E_2(\lambda t_i / t) - E_2(\lambda)\right],
\end{equation}
where $E_2(x) = \int_1^{\infty}t^{-2}e^{-xt}dt$ 
is the second-order exponential integral. $\lambda = t_i/t_{\mathrm{rec},i}$ is the ratio of the age of the universe at source ignition to the recombination time at that age. We set up the test with $n_{H,i} = 1.87\times 10^{-4}\,\mathrm{cm}^{-3}$ and $L_i = 7\times 10^{24}~$cm and using otherwise the same parameters as before. The result is shown in \autoref{fig:test4}, where $r_I(t)$ represents the comoving I-front radius, keeping in mind that $a(t_i)=1$.  For this test, we used the same cosmology as in \citetalias{c2ray}, namely $h = 0.7$, $\Omega_M=0.27$ and $\Omega_b = 0.043$.

\pyray{} again shows excellent agreement with the analytical result. While the effect of cosmic expansion is not evident at first sight, the analytical prediction without cosmology, \autoref{eq:rI1}, is also plotted for reference in the figure (green dotted line), and the difference is clearly visible. Again, results are almost as accurate when using a coarse time step.
\begin{figure*}[h]
    \centering
    \includegraphics[width=\linewidth]{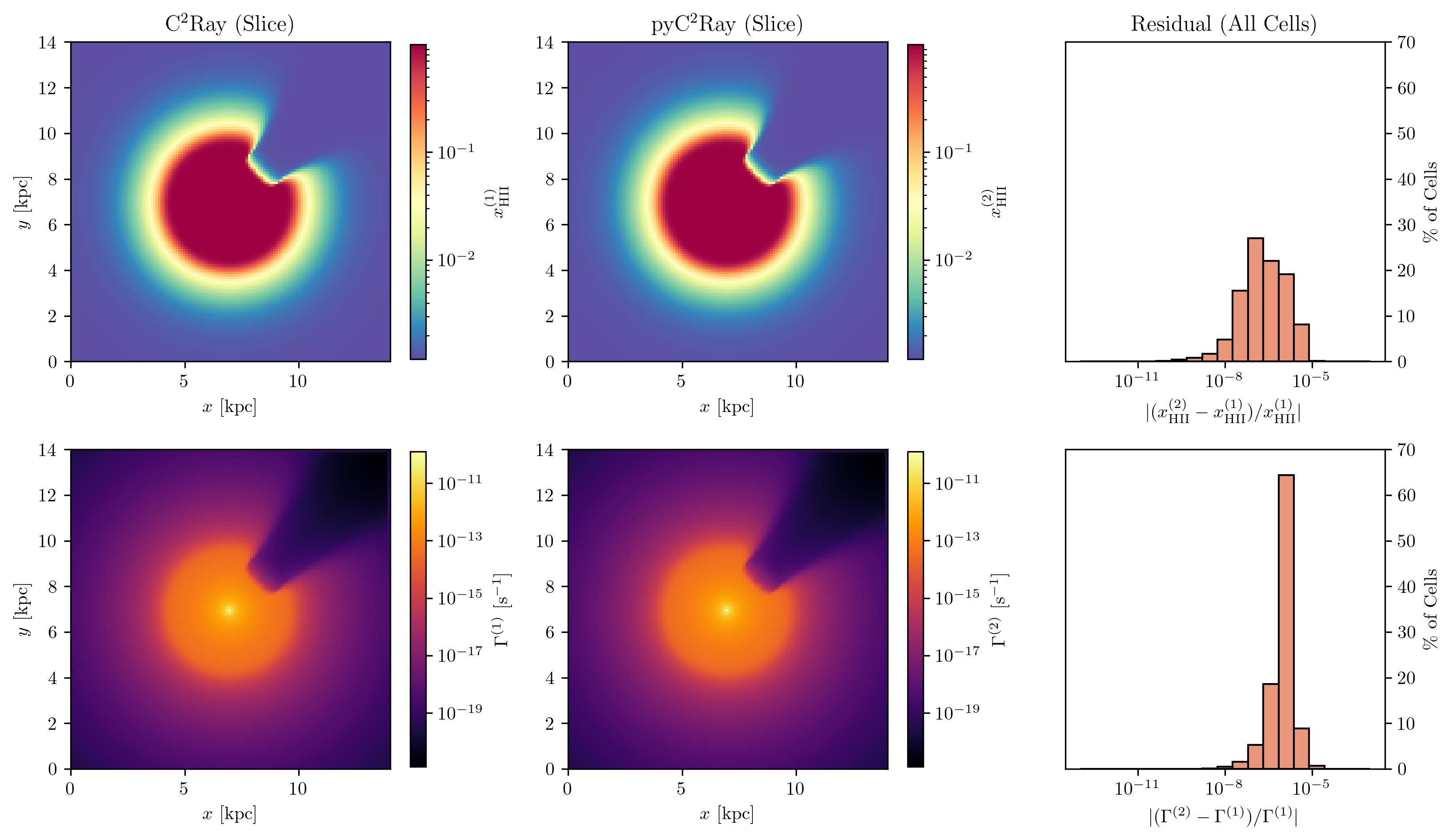}\vskip-4mm
    \caption{Result for Test 4 (I-Front Trapping in a Dense Clump and Formation of a Shadow). Shown are slices through the $z$-plane containing one ionizing source at the center and a dense clump of hydrogen diagonally offset from the source. The top row shows the ionized hydrogen fraction for \ray{} (left) and \pyray{} (middle), as well as the relative error between the two (right). The bottom row shows the same comparison for the photoionization rate.}
    \label{fig:shadow}
\end{figure*}
\subsubsection{Test 3: Expansion of Overlapping \hii{} regions around Multiple Black-Body Sources} \label{sec:test3}
Now we turn to the more realistic case of non-grey opacity and parameterize the cross section as $\sigma_{\nu} = \sigma_{0}(\nu/\nu_0)^{-\alpha}$, where $\nu_0$ is the ionization threshold frequency. The parameters of the power law are as in \citetalias{c2ray}, $\sigma_0 = 6.3\times 10^{18}$ cm$^{-2}$ and $\alpha=2.8$. We test how the ionization front is affected by the spectral characteristics of the sources. For harder spectra, where the energy peak is well above the ionization threshold, we expect wider ionization fronts, as the hard photons can penetrate deeper into the medium \citep{1998ppim.book.....S}. To test this and at the same time visualize how different \hii{} regions overlap, we place 5 black-body sources, each with total ionizing flux $\dot{N}_{\gamma} = 5\times 10^{48}$ but with different temperatures $T_{bb}$, in a dice-like pattern on the same $z$-plane. The box size is $L=14$ kpc, the mesh $128^3$ and the constant hydrogen density is $n_H = 10^{-3}$ cm$^{-3}$. We simulate for $t_{evo}=10\,\rm Myr$, with time step $\Delta t = 1\,\rm Myr$. \autoref{fig:multisource} shows cuts through the source plane of the final ionized hydrogen fraction $x_\mathrm{HII}$, for \pyray{} (top) and \ray (middle), along with the distribution of the \textit{absolute relative error}, $\left|(x_\mathrm{HII}^{\text{\pyray{}}} - x_\mathrm{HII}^{\text{\ray{}}}) / x_\mathrm{HII}^{\text{\ray{}}} \right|$, between the two coeval cubes (bottom panels). The leftmost column is the grey-opacity case as in the two previous tests, while the three remaining columns contain the results for $T_{bb}=\{5\times10^3, 5\times 10^4,1\times 10^5\}\, \rm K$.

Qualitatively, both \ray{} and \pyray{} reproduce the expected softness of ionization fronts for hot spectra, and the overlap of individual H II regions is also correctly modeled. The largest value for the relative error is on the order $10^{-4}$ in all cases, while the mean increases for harder spectra. Although relatively small, this error requires an explanation, as both codes should, in principle, produce equal results in the absence of unit conversion or floating point errors. In fact, an important technical difference between the two is the choice of numerical integration method used to pre-compute \autoref{eq:tables} as described in \S\ref{sec:computerate}. \pyray{} uses the standard \texttt{quad} wrapper of the \texttt{SciPy} package \citep{scipy2020}, which uses the adaptive quadrature method from the \texttt{QUADPACK} Fortran library. On the other hand, \ray{} uses a custom-written Romberg integration scheme. Both methods are valid choices, but they will inevitably yield slightly different results depending on the chosen resolution. We tested this by varying the frequency bins used by the Romberg method in \ray{} and found that the relative error between the two codes drops significantly as this number increases. We thus conclude that this technical difference is the most likely explanation for this result.

\begin{figure*}
    \centering
    \includegraphics[width=0.45\linewidth]{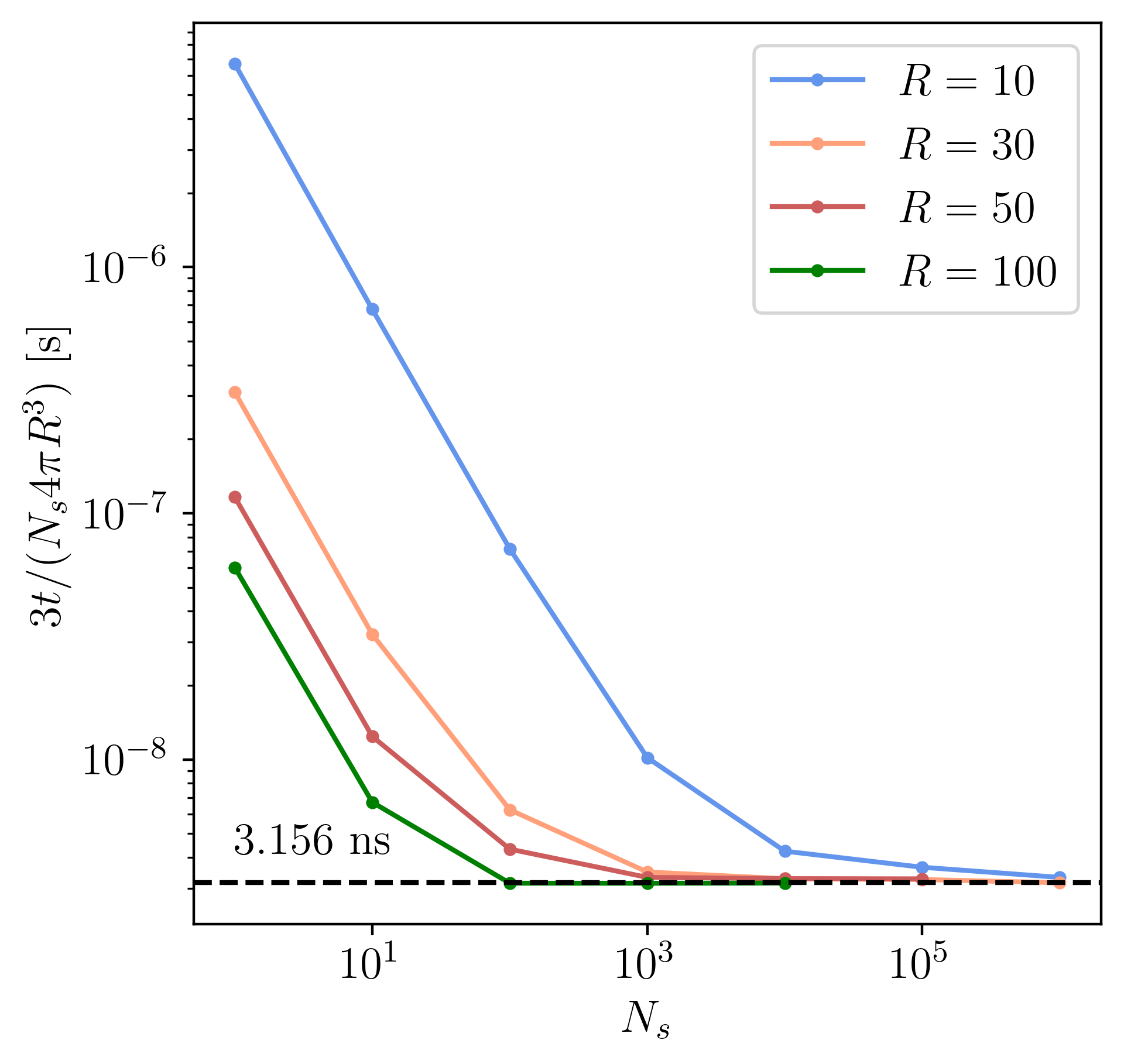}
    \includegraphics[width=0.45\linewidth]{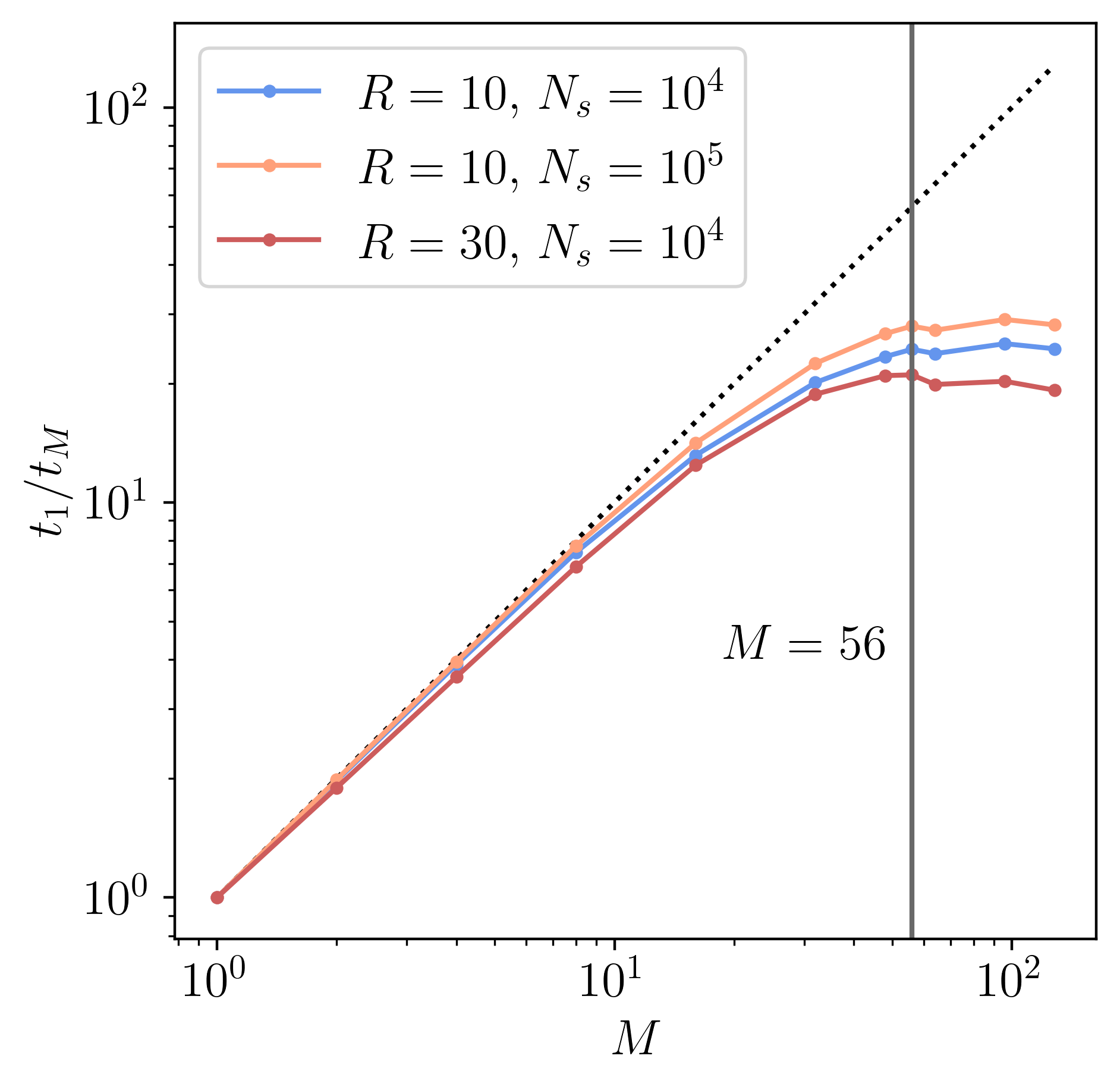}\vskip-4mm
    \caption{Scaling of the \texttt{ASORA} ray-tracing library. \emph{Left:} Computation time per source per voxel for an increasing number of sources $N_\mathrm{src}$ and different ray-tracing radii $R$. This time approaches a constant value as more sources are added and faster for larger $R$. \emph{Right:} Speedup in terms of the number of blocks $M$, given by $t_1/t_M$, where $t_1$ is the timing when a single block is used. The vertical black line marks $M=56$, corresponding to the number of SMs on the NVIDIA\textsuperscript{\textregistered} P100 GPU used in this benchmark.}\label{fig:scaling}
\end{figure*}

\subsubsection{Test 4: I-Front Trapping in a Dense Clump and Formation of a Shadow} \label{sec:test4}
Finally, to probe more specifically the ray-tracing method, we test for the formation of a shadow behind an overdense region. Correct modeling of shadows is one of the key advantages of ray-tracing over other techniques, making this an important check. In this test, the box size is $L=14$ kpc with mesh $128^3$ and a source with total ionizing flux $\dot{N}_{\gamma} = 10^{49}$ s$^{-1}$ is placed at its center. The hydrogen has a mean density $\bar{n}_{\rm H}=10^{-3}$ cm$^{-3}$, and a spherical overdense region of radius $r=8.75$ pc is placed on the same $z$-plane as the source, at a distance $d=2.01$ kpc diagonally from it. Within this region the density is $n_\mathrm{H}^{\star} = 6\,\bar{n}_{\rm H}$. The source has a black body temperature $T_{bb}=5\times 10^4$ K, and $t_{\rm evo}$ and $\Delta_t$ are as in \textit{Test 3}. The result is visualized in \autoref{fig:shadow}, where a cut through the source plane of the final ionized hydrogen fraction $x_\mathrm{HII}$ is shown on top and the photoionization rate $\Gamma$ below, for both codes along with the relative error as before. We want to point out that the fuzziness of the shadow is a feature of the short characteristic ray-tracing. The relative error is small again, and we believe it to be due to the choice of integration method used in the previous test. Interestingly, this error is larger by an order of magnitude at the edge of the overdense region. This is not so surprising, given that the overdensity is very optically thick and thus contains a large density gradient at its boundary. We noticed that the relative error is negative closer to the source, then positive, and then close to 0, reflecting the net photon flux conservation.
\begin{figure*}
    \centering
    \includegraphics[width=0.4\linewidth]{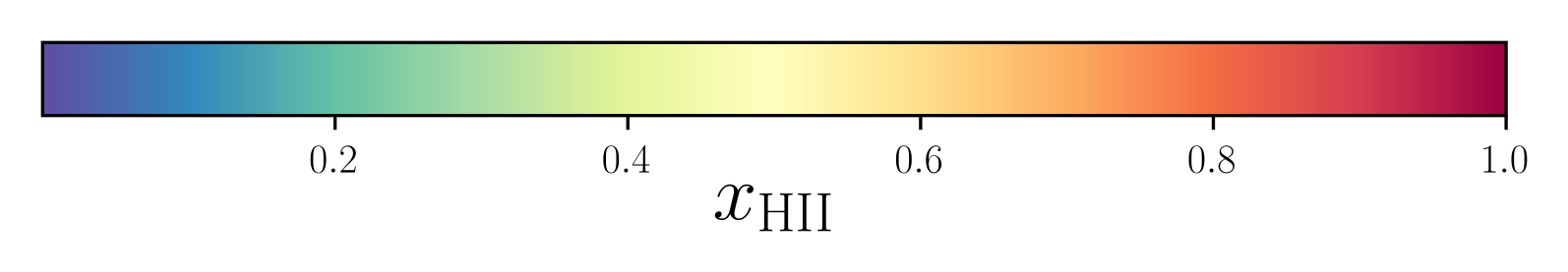}
    \includegraphics[width=0.88\linewidth]{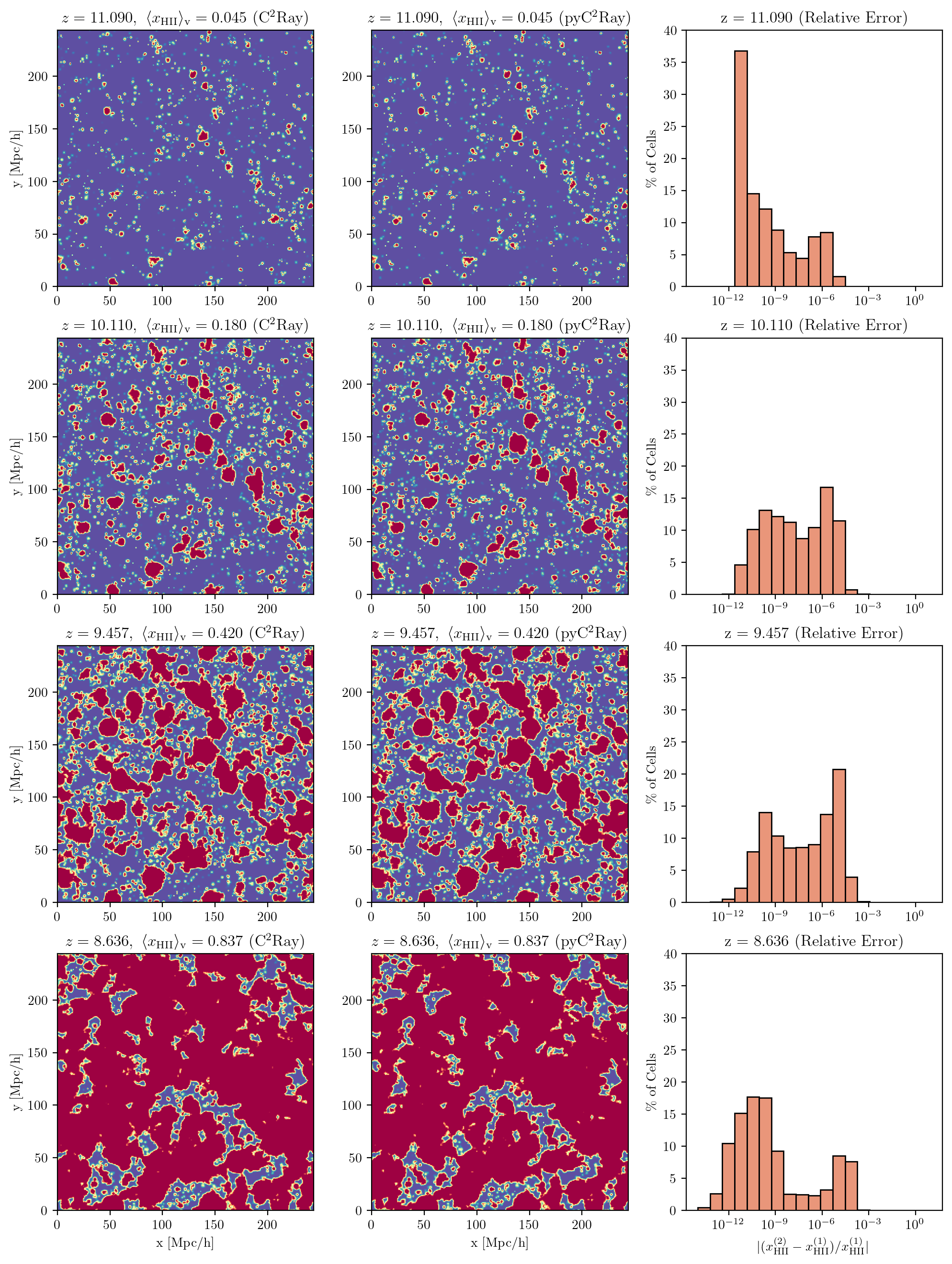}
    \caption{Results from of the $349\,\rm cMpc$ EoR test simulation. The left and middle columns show slices through the simulation domain for \ray{} and \pyray{}, respectively. The right column shows the distribution relative per-voxel error for the 250$^3$ grid. The simulation includes only dark matter halos masses with an efficiency factor $f_{\gamma} = 30$ and a maximal comoving photon radius $R = 15\,\rm cMpc$.}
    \label{fig:slice244}
\end{figure*}
\begin{figure*}
    \centering
    \includegraphics[width=0.45\linewidth]{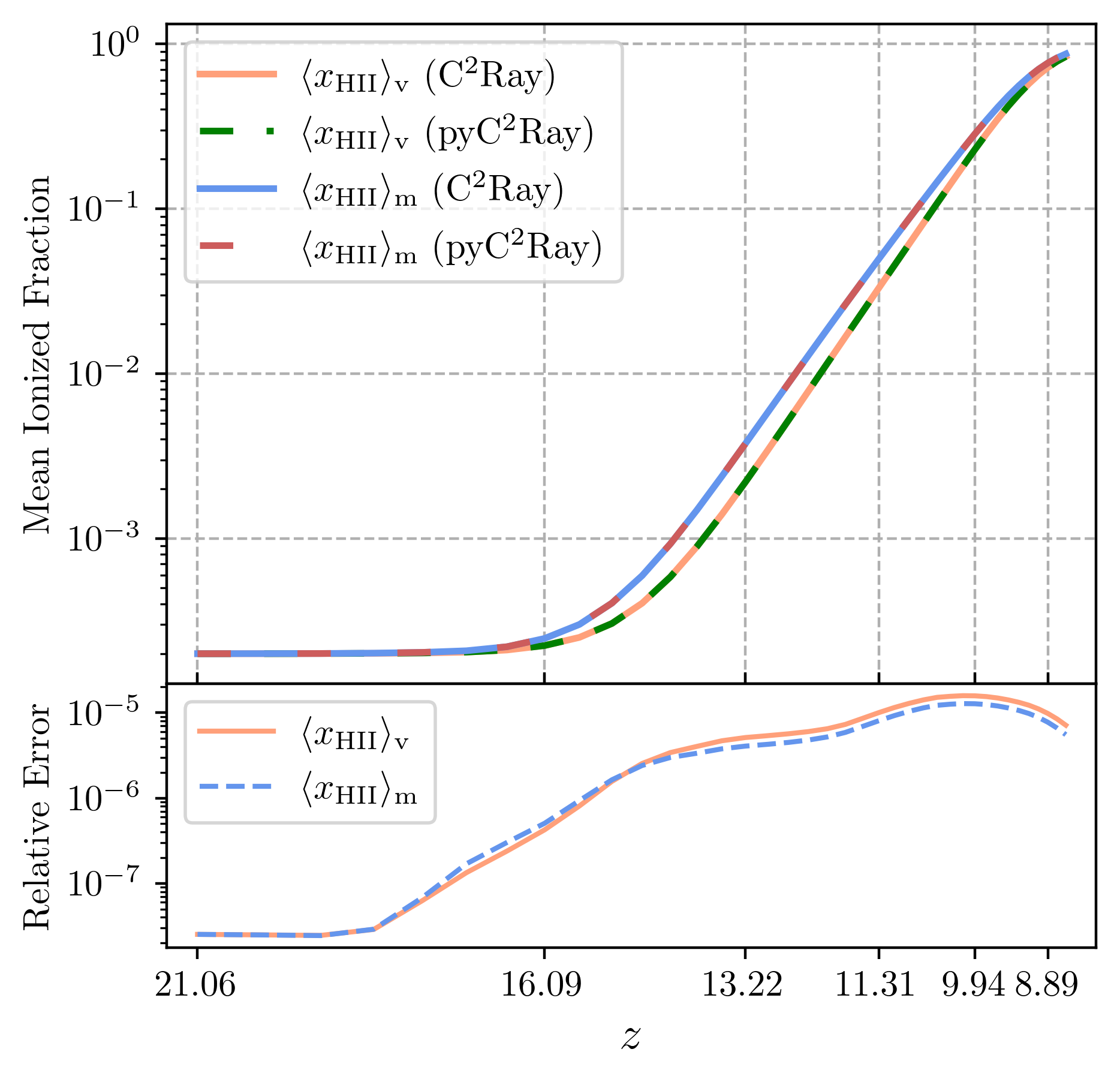}
    \includegraphics[width=0.45\linewidth]{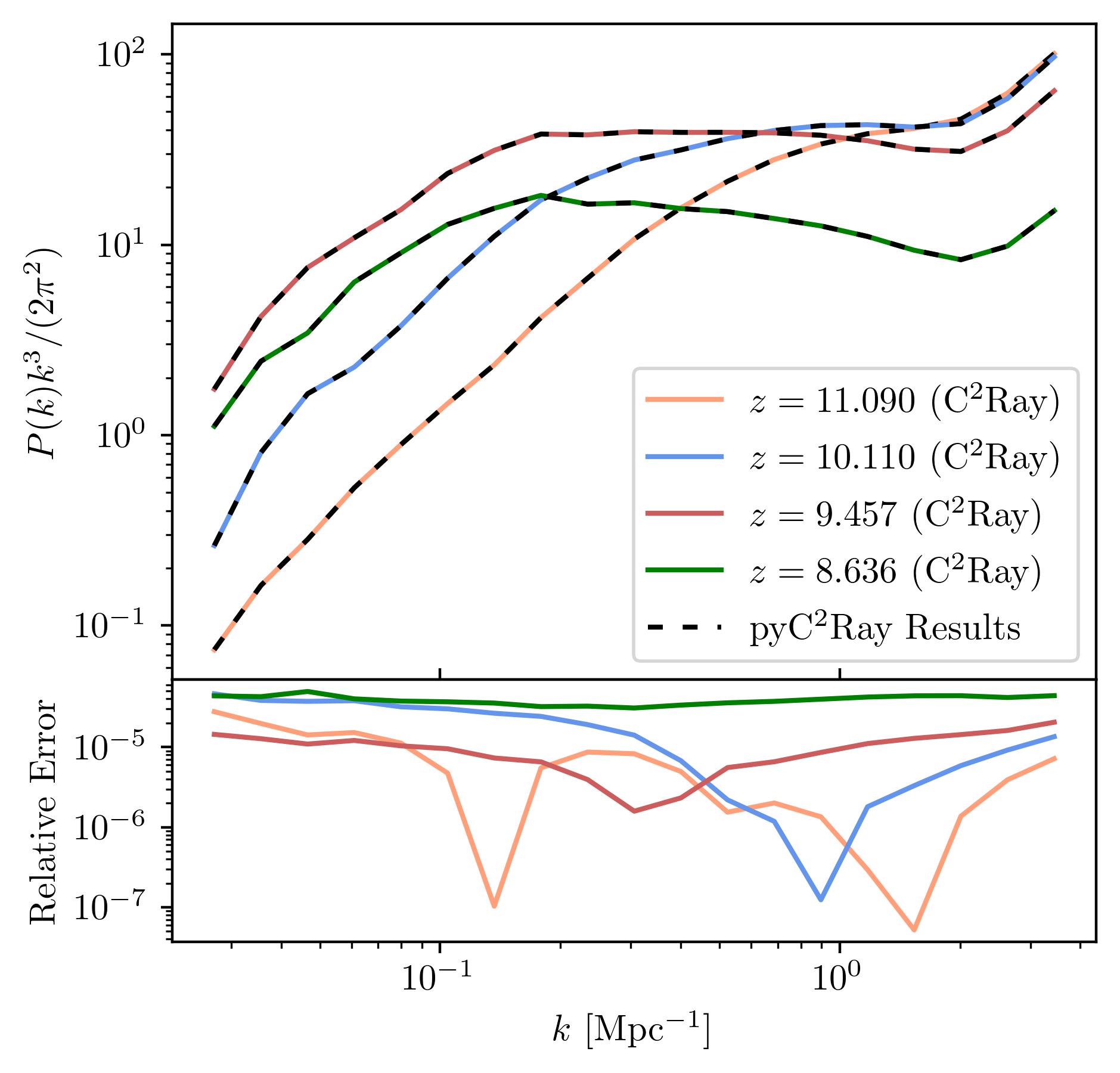}\vskip-4mm
    \caption{\textit{Left:} Comparison of reionization history from the 349 Mpc EoR test simulation, performed with \ray{} and \pyray{}. The top panel shows the evolution of the volume $\langle x^\mathrm{v}_\mathrm{HII} \rangle$ and mass-averaged $\langle x^\mathrm{m}_\mathrm{HII} \rangle$ fraction of ionized hydrogen over the redshift range $z\in [21.06,8.636]$ and the bottom panel the relative error between the two codes. \textit{Right:} Comparison of the 21-cm power spectra from the same simulation at different redshifts (indicated in the legend) reveals a consistent match between \ray{} and \pyray{}.}
    \label{fig:eorhist244}
\end{figure*}
\subsection{Performance Benchmark} \label{sec:performance_benchmark}
We now examine the performance of the new ray-tracing library more closely. All benchmarks in this section are performed on a size $N=250$ grid and run on one node of the Piz Daint\footnote{\url{https://www.cscs.ch/computers/piz-daint/}} computer at CSCS, containing in particular a single NVIDIA\textsuperscript{\textregistered} Tesla P100 GPU. First, we determine how the ray-tracing performance scales as more sources are added or the radius of ray-tracing per source increases. We expect the code to scale linearly with the number of sources $N_\mathrm{src}$ and as $\mathcal{O}(R^3)$ with the ray-tracing radius, $R= N_{mesh}\cdot R_{\rm max}/L_{\rm B} $, where $R_{\rm max}$ is the maximum radius for ray-tracing and $L_{\rm B}$ the box size, both in $\rm cMpc$ units. The benchmark is set up as follows. For $R=[10,30,50,100]$, the ray-tracing routine is called (on its own, without solving the chemistry afterward) on $N_\mathrm{src} = 10^a,\ a=0,\dots,6$ sources, and its run time is averaged over 10 executions. The left panel of \autoref{fig:scaling} shows the computation time per source per voxel,
\begin{equation}
    \Delta t(N_\mathrm{src},R) = \frac{t(N_\mathrm{src},R)}{\frac{4}{3}\pi R^3 N_\mathrm{src}},
\end{equation}
where $t(N_\mathrm{src},R)$ is the run time of the function running on $N_\mathrm{src}$ sources and computing $\Gamma$ in a spherical volume of radius $R$ (in voxel units) for each of them. With increasing $N_\mathrm{src}$, $\Delta t(N_\mathrm{src},R)$ approaches a constant value of about $3.156\,\rm ns$ on our system. Furthermore, this convergence is faster when the radius $R$ is larger. This implies that when few sources are present, overheads represent a non-negligible fraction of the execution time, even more so when the work per source (determined by $R$) is low. However, we can see that above $\sim 1000$ sources, the execution time is very close to its minimum, even for a relatively small RT radius. With few sources, the total amount of work is low and is not an expensive calculation. But typically, EoR simulations require $N_\mathrm{src} \gg 1000$. Our code runs in a regime where the work and not overheads dominate the performance of the code.

Next, we test how the code scales as the source batch size $M$ increases, corresponding to increasing the number of \texttt{CUDA} blocks dispatched to the device between global synchronizations. The right panel of \autoref{fig:scaling} presents the speedup $t_1 / t_M$ (where $t_M$ is the execution time using $M$ blocks) achieved in 3 cases; $(R=10,N_\mathrm{src}=10^4)$, $(R=10,N_\mathrm{src}=10^5)$ and $(R=30,N_\mathrm{src}=10^4)$ to see the impact of both the radius and total number of sources. This test is an analog of the "strong scaling" measurement typically performed on CPU cores. We observe that on our system, in all 3 cases, the code scales well up to $M\sim 32$ and does not gain any performance above $M\sim 50$, which seems to indicate that the sequential portion of the code prevents further scaling (analogously to Amdahl's law in CPU computing). This test, however, only gives a picture of the speedup achieved \emph{relative} to the single-block case for the whole program and hence does not indicate how good the occupancy of the GPU itself is. Detailed profiling using standard NVIDIA software has revealed that the number of registers per thread required by the ray-tracing kernel is likely a limiting factor that prevents the code from ever reaching maximum occupancy in its current state, even on GPUs with higher compute capability than the P100. Overcoming this limitation should be one of the main targets for future performance updates.

Two conclusions arise from this section: (1) The library is most optimized for use cases where many sources are present in the simulation, as is the case in EoR modeling. However, in cases where few sources are present, it will run optimally if the number of raytraced voxels is large. This may be the case when performing high-resolution radiative transfer simulations of smaller volumes, thus expanding the possible usage scenarios for \pyray{}. (2) A good value for the batch size $M$ will depend strongly on the system on which the code is run while simultaneously being limited by the available memory. This is because each block needs a cache space for the ray-tracing, the size of which scales with the grid, i.e., $\mathcal{O}(N^3)$.

\section{Running a Cosmological Reionization Simulation}\label{sec:running_cosmo}
The ultimate test for the updated code is to see whether it can reproduce the results of a simulation performed with the original \ray{} while at the same time achieving a gain in performance. Here, we post-process a $(349 \mathrm{Mpc})^3$ volume $N$-body simulation run with $4000^3$ dark matter particles, which models the formation of high-redshift structures. These $N$-body simulations used the code \texttt{CUBEP$^3$M} \citep{Harnois-Deraps2013}\footnote{\url{https://github.com/jharno/cubep3m}}, which has an on-the-fly halo finder, providing halo catalogs at each redshift snapshot using the spherical overdensity method \citep[see][for more detail]{watson2013halo}. The $N$-body dark matter particles and the halo catalog are then gridded, with an SPH-like smoothing technique, onto a regular grid of size $N_{\rm mesh}=250^3$ that is later used as inputs for the RT simulation. This simulation resolves dark matter haloes with mass $M_{\rm halo} \geq 10^9 {\rm M_{\odot}}$. This simulation contains approximately $10^7$ sources toward the end of reionization. See \citet{Dixon2016Largescale} and \citet{giri2018Bubble} for more detailed descriptions.

We follow the same source model presented in previous work \citep[e.g.][]{Iliev2014Howbig, Bianco21Inhomo} that assumes a linear relation between the emissivity and the mass of the hosting dark matter halo. In this model, the grand total of ionizing photons, $\dot{N}_{\gamma} $, produced by a source residing in dark matter halo mass $M_{\rm halo}$ is
\begin{equation}
    \dot{N}_{\gamma} = f_\gamma\,\frac{M_{\rm halo}\,\Omega_{\rm b}}{\Omega_{\rm M}\, m_p\, t_s },
\end{equation}
where the efficiency factor $f_{\rm \gamma} = 30$ and the source lifetime $t_s\approx 10\,{\rm Myr}$ is taken to be the time difference between the simulation snapshots. Two time steps are performed for each redshift interval.
Here, we choose an extreme value for the efficiency factor to speed up the reionization process so we could run \ray{} in a reasonable amount of time and computational resources. 
We should note that reionization ends quite early compared to more realistic models in \citet{Dixon2016Largescale} and \citet{giri2019neutral} produced using \ray{}—however, the outcomes of the comparison hold for any source model. 

In \autoref{fig:slice244}, we show slices of the simulated ionized fraction, $x_\mathrm{HII}$, comparing \ray{} (left column) and \pyray{} (middle column) at redshift $z=11.090,\,10.110,\,9.457$ and $8.636$, corresponding to a volume-averaged ionized fraction $\left<x_\mathrm{HII} \right>=0.045,\,0.180,\,0.420$ and $0.837$. We show the relative error in the right column of the same figure for each redshift. At high redshift, the error distribution is mostly centered at $10^{-6}$, similar to what we show in \S~\ref{sec:test3} and \ref{sec:test4}. While from $z \sim 10$, it shows two peaks with the distribution transitioning from $10^{-4.5}$ to $10^{-10}$. The double-peaked feature of the error distribution is visible from the moment the source contribution becomes substantial. This indicates that the error distribution is initially associated with the precision error in the vast neutral field while later with the growing ionized regions. In the left panels of \autoref{fig:eorhist244}, we calculate the volume- and mass-averaged ionized fraction, $\left<x_\mathrm{HII} \right>_{\rm v}$ and $\left<x_\mathrm{HII} \right>_{\rm m}$, against redshift. With solid lines, we indicate the results obtained with \ray{}, while in dashed lines, the one with \pyray{}. Similar to what we show in the previous paragraph, on average, the relative error is at least five orders of magnitude smaller, $\sim 10^{-5}$, compared to the dynamic range of the ionized field, making the difference indiscernible. Notice that we show the result to $z=8.575$ when the IGM is about $86\%$ ionized. However, at this reionization epoch, the simulation has approximately $\sim 1.5\times 10^{6}$ sources, and \ray{} starts to become computationally demanding.

Radio experiments, such as HERA, LOFAR, and MWA, aim to observe the spatial distribution $\pmb{r}$ of the differential brightness temperature $\delta T_\mathrm{b} (\pmb{r},z)$ corresponding to the 21-cm signal. This quantity can be given as \cite[e.g.][]{pritchard201221},
\begin{equation}
\begin{aligned}
    \delta T_{b}(\pmb{r},z) \approx 27~\mathrm{mK}&\left(\frac{0.15}{\Omega_{M}h^{2}}\frac{1+z}{10}\right)^{\frac{1}{2}} \left(\frac{\Omega_{ b}h^{2}}{0.023} \right) \\ 
    &\times [1-x_\mathrm{HII}(\pmb{r},z)][1+\delta_{\rm b}(\pmb{r},z)],
  \label{eq:dTb formula}
\end{aligned}
\end{equation}
where $x_\mathrm{HII}$ and $\delta_\mathrm{b}$ are ionization hydrogen fraction and baryon overdensity, respectively. We should note that we have assumed a spin temperature to be saturated and ignore the impact of redshift-space distortion. We refer the interested readers to \citet{ross2021redshift} for exploration of both these aspects in simulations with \ray{}.
We compute $\delta T_\mathrm{b} (\pmb{r},z)$ and subsequently the power spectrum using reionization simulation snapshots with our data analysis software, \texttt{Tools21cm}\footnote{\url{https://github.com/sambit-giri/tools21cm}} \citep{giri2020tools21cm}. In the top-right panel of \autoref{fig:eorhist244}, we present the 21-cm power spectrum at various redshifts. We observe a precise agreement between the results obtained from \pyray{} and \ray{}, also evident from the relative error in the bottom-right panel, demonstrating that these upgrades can accurately replicate the spatial distribution of the 21-cm signal.

The simulation with \pyray{} cost $2.5$ GPU-hours on our single-GPU system, while the comparison run, with the \texttt{Fortran90} CPU version of \ray{}, computed on 128 cores for a total of $13,824$ core-hours. While GPU hours are, in general, more expensive than core hours, the observed speedup is so large that \pyray{} is significantly cheaper to run than the original code by a factor of $\sim 100$, depending on the computing center, which was part of the motivation behind this update. In \ref{sec:carbon}, we illustrate further the computational advantage of porting algorithms to GPU.

\section{Summary and Conclusions}
The main challenge in simulating the cosmic Epoch of Reionization is that we must concurrently simulate a large volume of the order of the $\rm Gpc$ scale while resolving compact and dense cosmic structures. These requirements make Radiative Transfer (RT) simulations extremely computationally expensive and demanding. For this reason, most RT codes are implemented with programming languages suited for scientific computing, such as \texttt{Fortran90} or \texttt{C/C++}. However, this makes any changes or regular updates to the code cumbersome for new users, as any slight modification requires frequent recompilation and debugging. Moreover, relatively little effort has been made to make ray-tracing algorithms for reionization simulations computationally efficient and functional on general-purpose graphic process units (GPU).

Therefore, this paper introduces \pyray{}, a \texttt{Python} wrapped updated version of the extensively used \ray{} RT code for cosmic reionization simulations. In particular, we present the newly developed Accelerated Short-characteristics Octhaedral RAy-tracing algorithm, \texttt{ASORA}, that utilizes GPU architectures to achieve drastic speedup in fully numerical RT simulations. 

In \S~\ref{sec:c2ray}, we recap the differential equation solved during a cosmological reionization simulation. In \S~\ref{sec:c2raysummary}, we summarize the well-established time-averaged method that solves the chemistry equation in \ray{}, \autoref{eq:chemistry}, allowing the solution to be integrated on a larger time-step compared to the reionization time scales, otherwise required by a more direct approach. In \S~\ref{sec:computerate}, we explain in detail the necessity for an efficient ray-tracing method for our code. With \autoref{eq:gamma} and \ref{eq:sugg}, we highlight the core and most computationally expensive operation in RT algorithms, which consists of computing the column density and, thus, the optical depth for each voxel, that ultimately quantifies the number of ionizing photons that are absorbed by a cell along the ray. The combination of the time-averaged and short-characteristics methods are the distinguishing features of the \ray{} code. In \autoref{fig:c2ray_flowchart}, we summarize the algorithm for both the \ray{} and \pyray{} methods presented here.

In \S~\ref{sec:ray-tracing_c2ray}, we remind the reader of the short-characteristic approach of \ray{} inherited by \pyray{}. In \S~\ref{sec:CPU_parallelization}, we describe the existing CPU parallelization of the current version of \ray{}, which consists of splitting the source input list into equal parts for each \texttt{MPI} processor. For each rank, 8 \texttt{OpenMP} threads, corresponding to the number of independent domains around each source, compute the $\rm HI$ column density. This parallelization strategy is not optimal for GPU architectures. Therefore, in \S~\ref{sec:GPU_parallelization} we propose a new interpolation approach for the \ray{} RT algorithm specifically designed for GPUs. The \texttt{ASORA} interpolation scheme comes from the physical intuition that the radiation propagates as an outward wavefront around a source. This new approach changes the domain decomposition to an interpolation between concentric surfaces of an octahedron centered around the source as illustrated by \autoref{fig:octa}. From a technical perspective, in \pyray{}, we keep the same \texttt{MPI} source distribution, as presented in \S~\ref{sec:CPU_parallelization}, and instead replace the \texttt{OpenMP} domain decomposition with the \texttt{ASORA} method. 

The update also includes the conversion to \texttt{Python} of the non-time-consuming subroutines of \ray{}. In \S~\ref{sec:pyton_wrapping}, we mention how the use of commonly used libraries, such as \texttt{Numpy}, \texttt{Scipy} and \texttt{Astropy} can be easily included according to the user's need. Moreover, the \pyray{} user interface makes it easier to employ other codes that have also been \texttt{Python}-wrapped. For instance, we can easily incorporate in \pyray{} photo-ionization rates from other spectral energy distributions calculated with a population synthesis code such as \texttt{PEGASE-2} \citep{Pegase2011} or a different chemistry solver such \texttt{GRACKLE} \citep{Grackle2017}.

In \S~\ref{sec:results}, we show \pyray{} results on a series of standard RT tests. In \S~\ref{sec:test1} and \ref{sec:test2}, we demonstrate that \pyray{} agrees with the analytical solutions of the ionization front size, $r_{\rm I}$, for the single sources in a static and expanding lattice. To test that the conversion to \texttt{Pyhton} of the non-time-critical subroutines was successful and does not introduce substantial differences, in \S~\ref{sec:test3}, we test the results on overlapping $\rm HII$ regions for sources with different black body spectra. In \S~\ref{sec:test4}, we probe the formation of a shadow behind an overdense region, a standard test for ray tracing methods. 

In \S~\ref{sec:performance_benchmark}, we examine the performance of the new ray-tracing methods accomplished on the Piz Daint cluster at the Swiss National Supercomputing Centre (CSCS) equipped with an NVIDIA\textsuperscript{\textregistered} Tesla P100 GPU. Our main finding is that the \texttt{ASORA} RT computing time grows linearly with the increasing number of sources, $N_\mathrm{src}$, and in cubic fashion with respect to the maximum radius for ray tracing, so $R^3$, i.e., distance is given in a number of voxels. In the case of the Tesla P100 GPU, the computing time per source per voxel within the ray-tracing distance saturates with value $3.156\,\rm ns$ when $N_\mathrm{src}>10^5$. This study allows the user to quantify the computing time and cost of a future simulation run with \pyray{}. If we consider a cosmological simulation with 68 redshift steps, each with 2-time steps, ray-tracing radius $R=11$ (grid units) and approximately $N_\mathrm{src}\approx4\times10^6$ sources. We can run the entire simulation from $z=21$ to $8.5$ with a total of $\sim2.75$ GPU-h, corresponding to the cost obtained in the cosmological example presented in \S\ref{sec:running_cosmo}. Secondly, the method scales strongly with the batch size up to $\sim 32$ on our system, suggesting that the GPU occupancy is not yet optimal, an issue that may be addressed in future updates. We estimate that running a reionization simulation on the same volume down to $z\sim6$, where $N_\mathrm{src}=1.5\times10^7$, would cost approximately 10.3 GPU-h.

Finally, in \S~\ref{sec:running_cosmo}, we compare \pyray{} and \ray{} on an actual cosmological simulation. We demonstrate that the differences within the same simulation are negligible with an absolute-relative error between $10^{-4}$ and $10^{-12}$ on the $\rm HII$ field, while both mass- and volume-averaged ionized fractions and the power spectra accumulate an error that stays below the order of $<10^{-5}$. As mentioned in the previous paragraph, the computational cost for this simulation was $2.5$ GPU-hours, while the same simulation run on $128$ cores with \ray{} took $\sim14\rm k$ core-hours.  Another way to describe the gain in performance is to consider the monetary cost of running these simulations. The cost of running a code on a GPU or CPU cluster varies based on the electricity consumption and other indirect expenses assessed by the high-performance computer facility. Nowadays, one GPU-hour can cost on average $0.8\,\rm Euros$, while one core-hour can be $0.01\,\rm Euros$. Therefore, with these reference fees the simulation presented \S~\ref{sec:running_cosmo} would have cost $2\,\rm Euros$ if run with \pyray{} instead of $138.25\,\rm Euros$ with \ray{}.

With this work, we demonstrate that \pyray{} achieves the same result as \ray{} for a cosmological EoR simulation, but with a computing cost and time two orders of magnitude lower than the original code, confirming the motivation behind this modernization of \ray{}. In principle, \pyray{} is not limited by the volume size or the mass resolution but rather by the spatial resolution, $N$, and the number of sources, $N_{src}$. The \texttt{ASORA} raytracing algorithm needs to store $M$ copies of the entire double precision grid data directly on the GPU, where $M$ is the source batch size. Therefore, the current limiting factor is the available memory on the GPU, as it is generally desirable to have $M \gtrsim 20$ to achieve optimal GPU occupancy.  For instance, the NVIDIA\textsuperscript{\textregistered} P100 has 64 GB of memory; we can, in principle, simulate a $1024^3$ mesh grid but are then limited to $M<8$, which is below the optimal regime. We plan to address this issue by reducing the per-source memory requirement in those cases where the ray-tracing radius is significantly smaller than the whole box and the current implementation is needlessly memory-hungry. In this update, we focused on the simplest simulation setup, namely, no photo-heating and only photo-ionization for hydrogen chemistry. As mentioned, \ray{} has been extended to also include helium \citep{helium} and X-ray heating \citep{ross2017simulating}, and has also been used as a module in a hydrodynamic simulation to follow the evolution of an \hii{} region in the interstellar medium (ISM), see \cite{Henney2011} and \cite{Medina2014}. We aim to gradually include these features and extensions in \pyray{} now that the groundwork has been laid.

%
\section*{Acknowledgements}
The authors would like to thank  Emma Tolley, Shreyam Krishna and Chris Finlay for their feedback and useful discussions, as well as Hannah Ross, Jean-Guillaume Piccinali, Andreas Fink and Dmitry Alexeev for their help on the technical aspects of the GPU implementation. MB acknowledges the financial support from the Swiss National Science Foundation (SNSF) under the Sinergia Astrosignals grant (CRSII5\_193826). PH acknowledges access to Piz Daint at the Swiss National Supercomputing Centre, Switzerland, under the SKA's share with the project ID sk015. This work has been done as part of the SKACH consortium through funding from SERI. GM’s research is supported by the Swedish Research Council project grant 2020-04691\_VR. We also acknowledge the allocation of computing resources provided by the National Academic Infrastructure for Supercomputing in Sweden (NAISS) at the PDC Center for High-Performance Computing, KTH Royal Institute of Technology, partially funded by the Swedish Research Council through grant agreement no. 2022-06725. 

The image processing tools operated on our data were performed with the help of \texttt{NumPy} and \texttt{SciPy} packages. All plots were created with \texttt{mathplotlib} \citep{Hunter2007}, and the illustration in \autoref{fig:octa} was made using \href{https://www.blender.org/}{Blender}.

\appendix
\section{ASORA Implementation Details}\label{sec:appendix}
\begin{table*}
\centering
    \caption{Summary of the carbon footprint consumption of the cosmological simulation presented in this paper if both runs were performed in Switzerland.}
    \begin{tabular}{|l|c|c|c|c|c|}
        \hline
        Model             & CO$\rm_2$ emission [$kg$] & Energy consumption [$kWh$] & Car drive $[km]$ & CO$\rm_2$ absorption [$yr$] \\ \hline
        NVIDIA Tesla P100 & $0.02$                    & $1.07$                     & $0.07$           & $1.12$                      \\ \hline
        AMD EPYC Zen 2    & $1.22$                    & $105.88$                   & $6.97$           & $110.5$                     \\ \hline
    \end{tabular}\label{tab:co2e}
\end{table*} 

\begin{figure}[h]
    \centering
    \includegraphics[width=\linewidth]{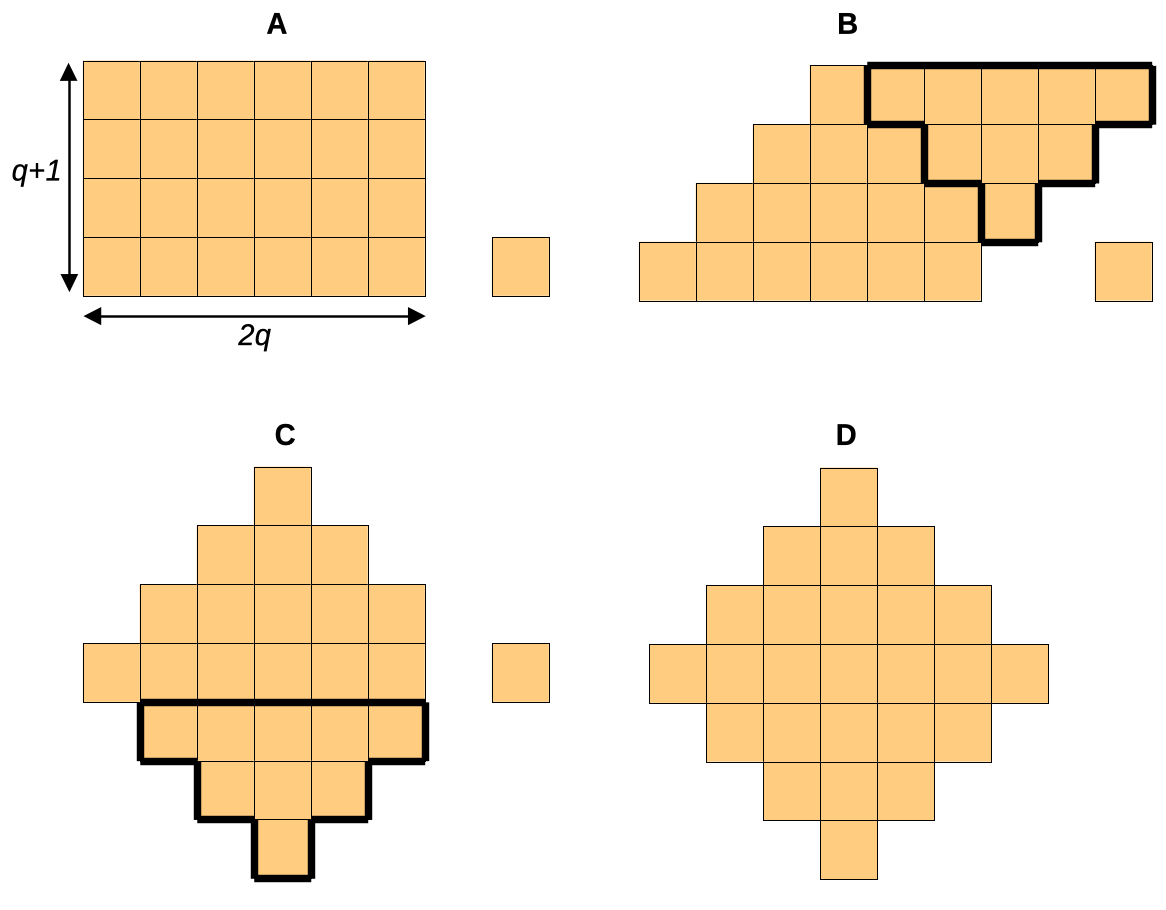}
    \caption{Schematic representation of the mapping of 1D indices $0,\dots,2q(q+1)$ to the 3D grid positions $(i,j,k)$ of voxels in the top part of the $q=3$ shell. The $(i,j)$ mapping is a combination of a shear (B) and a translation (C), and the $k$ coordinate is determined directly from $(i,j)$ as described in the text.}
    \label{fig:thread_mapping}
\end{figure}

Here, we briefly discuss how the \texttt{ASORA} method is implemented in \texttt{C++/CUDA}. As detailed in the paper, each block is assigned to a single source and owns a dedicated memory space to store the values of the column densities of voxels to be used as interpolants in upcoming tasks. Each task $S_q$ comprises the $|S_q| = 4q^2 + 2$ grid voxels belonging to an octahedral shell as illustrated in \autoref{fig:octa}. Threads within a block are labeled by 1D indices $x=0,\dots, N$, where $N$ is the block size. Labeling the voxels in the shell by $s=0,\dots,|S_q|$, all voxels can be treated if the threads iterate $\sim |S_q|/N$ times. It then remains to map the 1D indices $s$ to the actual 3D grid positions $(i,j,k)$ of the voxels within the shell. We use the following mapping: separate the octahedron into a "top" part containing all $k\geq k_s$ planes, where $k_s$ is the source plane, and a "bottom" part containing the rest. For the top part, which contains $2q(q+1) + 1$ voxels in total, the $k$ index of any voxel can be found from its $i,j$ indices through $k = k_s + q - (|i-i_s| + |j-j_s|)$. To find $i,j$, we follow the procedure illustrated in \autoref{fig:thread_mapping}: map $s=1,\dots,2q(q+1)$ to Cartesian 2D coordinates $(a,b)$ as in (A) and apply a shear matrix $(a,b)\rightarrow (a',b')$ to obtain (B). Apply a translation on the subset of those with $a+2b > 2q$ (C) and finally map the remaining voxel $s=0$ to $(i,j) = (i_s + q,j_s)$ to obtain the full squashed top part of the octahedron (D). The same procedure is applied to the lower part, with some slight modifications, as this does not include the source plane and so contains fewer voxels in total ($2q^2 - 1$). For further details, we refer the reader to the source code.
\label{sec:cuda}

A last key point to address is that since \ray{} uses periodic boundary conditions, it is important to impose a further constraint on the indices $(i,j,k)$ of the voxels that are allowed to avoid race conditions on coordinates that map to the same voxel under periodicity. The simulation domain is cubic, so this constraint is satisfied if we impose that no voxel can be farther away from the source than the edges of the grid, translated under periodicity.
On an odd mesh ($N$ odd), this means only considering voxels at most a grid distance $N/2$ away from the source on either side. On an even mesh, a convention must be chosen, and in line with the original \ray{} code, we impose that the maximum distance in each dimension is $N/2$ on the negative and $N/2-1$ on the positive side of the source.

\section{Carbon footprint of cosmological simulations}\label{sec:carbon}
Numerical simulations for cosmological and astrophysics applications often require immense computational power and extensive data processing, and therefore, their energy demands can be substantial. The environmental impact is often underappreciated and sometimes disregarded. Given the escalating concern over climate change, we want to present the ecological advantage of moving to GPU-based algorithms. 

We employed \texttt{Green Algorithm}\footnote{\url{www.green-algorithms.org}} to estimate the carbon consumption of the cosmological simulations presented in \S\ref{sec:running_cosmo} and compare the run with \pyray{} and \ray{}. As we mentioned in \S\ref{sec:performance_benchmark}, the cosmological EoR simulation presented in this paper run with \pyray{} was performed in 2 hours and 30min on 1 GPU NVIDIA\textsuperscript{\textregistered} Tesla P100, drawing 1.07 kWh. Based in Switzerland, this has a carbon emission (CO$\rm_2$e) of 12.31 g. This corresponds to the CO$\rm_2$ consumption of driving a car for 70 meters or 0.02$\%$ of the consumption of the Paris-London flight. Based in Sweden, the same simulation runs with \ray{} on 128 AMD EPYC Zen 2 CPUs. The cluster draws 105.88 kWh and has a 600.33 g CO$\rm_2$e, corresponding to the consumption of a car drive for 3.43 Km or the $1\%$ consumption of the Paris-London travel by plane. A mature tree sequesters on average 0.92 g of CO$\rm_2$ per month \citep{Lannelongue2021green}. Based on this estimation, the cosmological run performed, with \ray{}, would have consumed what one single tree sequester from the atmosphere in approximately 54 years. Meanwhile, the same simulation run with \pyray{} would take about one year. In \autoref{tab:co2e}, we compare the simulations CO$\rm_2$ consumption if both runs were performed in Switzerland.

While this analysis highlights the environmental footprint of cosmological simulations, its purpose is not to evoke shame or guilt. Rather, it serves as a reminder of the tangible costs of these essential scientific endeavors. Moreover, we did not consider using renewable energy sources and the potential impact reduction of HPC clusters using renewable energy. We aim to highlight the differences in energy consumption between simulation approaches.

\bibliographystyle{elsarticle-harv} 
\bibliography{bibliography}

\end{document}